\documentclass[12pt, aps, unsortedaddress]{article}
\usepackage[pdftex]{graphicx}
\usepackage{epsfig}
\begin{document}

\title{The standard model of spin injection}

\author{\vspace{2cm} \\ Jaroslav Fabian \\
Institute for Theoretical Physics\\
 University of Regensburg \\
  D-93040 Regensburg, Germany
     \vspace{0.5cm} \\ \&   \vspace{0.5cm} \\
Igor \v{Z}uti\'{c} \\
Department of Physics \\
State University of New York at Buffalo \\ Buffalo, NY 14260, USA}

\date{}

\maketitle

\newpage

\tableofcontents

\newpage

\section{Introduction}

The generation of nonequilibrium electron spin, as well as the
nonequilibrium spin itself, in electronic materials (metals and
semiconductors), is called {\it spin accumulation}.\footnote{By spin
in spin injection is meant a spin ensemble, rather than an
individual electron spin.} The most important techniques for spin
accumulation are electrical spin injection, optical spin
orientation, and spin resonance. By electrical spin injection, or
simply {\it spin injection}, we mean spin accumulation by injecting
spin-polarized electrons from one material to another, by electric
current. The source material could be a ferromagnetic metal, for
example Fe, in which there is a difference in the densities of spin
up and spin down electrons. Such a difference is characterized by a
spin {\it polarization}. In the ferromagnet the spin polarization
exists in equilibrium. In contrast, if electrons from the
ferromagnet are injected into a nominally nonmagnetic metal, say,
Al, the resulting spin polarization in Al is a nonequilibrium one:
spin accumulates in Al. Another possibility is an electrical spin
injection between two nonmagnetic materials, say Al and Cu. If one
of the materials has a nonequilibrium spin, electric current can
lead to spin accumulation in the other material. Electrical spin
injection is the main topic of these lecture notes.

The two other techniques for spin accumulation historically
preceded spin injection. Optical orientation is a process of
generating nonequilibrium spin optically, by exposing the material to
a circularly polarized light. The angular momentum of the photons is
transferred to the electron spin. Optical orientation is most
effective in direct band semiconductors such as GaAs. The
historically first technique for investigation nonequilibrium spin
has been electron spin resonance. Application of a magnetic field
splits the spin up and spin down electron states (Zeeman splitting)
with a corresponding equilibrium spin polarization. A microwave
radiation\footnote{Microwave photons have energies matching the
electron Zeeman splitting which is typically  $0.01 - 1$ meV's, in
fields of order tesla. Radio waves are typically used for nuclear
spin resonance.} can induce transitions between the spin-split
states, generating nonequilibrium spin. The spin resonance technique
has been used in metals and semiconductors. There are other ways to
generate spin accumulation, typically much less efficient as with
the three ways mentioned above. One example is the spin Hall effect,
in which electric current leads to a separation of spin up and spin down
electrons at the edges parallel to the current flow. Another
possibility is to first accumulate nuclear spin in the lattice ions;
electron spins can be then polarized via the hyperfine interaction.

The standard model of spin injection originates from the proposal of
Aronov \cite{Aronov1976:JETPL} who suggested the possibility of electrical
spin injection from a ferromagnetic to a nonmagnetic conductor. The
thermodynamics of spin injection has been developed by Johnson and
Silsbee, who also formulated a drift-diffusion transport model for
spin transport across ferromagnet/nonmagnet (F/N) interfaces
\cite{Johnson1987:PRB, Johnson1988:PRBa}. This model has been shown
to be essentially equivalent to the standard model as presented here
\cite{Zutic2004:RMP, Fabian2007:APS}. The theory of spin
injection was further developed in \cite{vanSon1987:PRL,%
Valet1993:PRB, Fert2001:PRB, Hershfield1997:PRB, Schmidt2000:PRB,%
Fabian2002:PRB, Zutic2002:PRL, Jedema2003:PRB, Rashba2000:PRB, Rashba2002:EPJ,%
Vignale2003:SSC, Jonker2003:MRS, Takahashi2003:PRB, Fert2007:IEEE,
Zutic2006:PRL}. In particular the presentation of Rashba
\cite{Rashba2000:PRB, Rashba2002:EPJ} has inspired the formulation
of the standard model of spin injection in the reviews
\cite{Zutic2004:RMP, Fabian2007:APS} which these lecture notes
follow and extend. These reviews should be consulted for original references
and examples of experimental results.

\section{Simple model of spin injection}

Perhaps the simplest model of spin injection considers a steady flow of a
spin-polarized electric current from a ferromagnet to a nonmagnetic
conductor. The ferromagnet has an electron spin polarization $P_0$;
for the present purposes $P_0$ is the relative difference between
the ``relevant'' densities of spin up and spin down electrons. More
specific definitions of the term are given later. In a typical
ferromagnetic metal $P_0$ is 10--50\%. In nonmagnetic metals the
spin polarization at equilibrium vanishes.

Calling the ferromagnetic conductor $F$ and the nonmagnetic
conductor $N$, we have a simple F/N junction. We wish to answer the
following question:

\vspace{0.2cm} {\it Given the equilibrium spin polarization $P_0$ in
the ferromagnet, what is the spin accumulation in the nonmagnetic
conductor if electric current $j$ flows through the junction?}
\vspace{0.2cm}

In order to answer this question, we need to know how much spin per
unit time arrives from $F$ to $N$. The simplest answer would be
$j_{s0}/(-e)$, where the spin current
\begin{equation} \label{eq:Aronov1}
j_{s0} = P_0 j,
\end{equation}
as the spins are attached to the electrons flowing through the
interface. We can take this value as a very rough estimate of what
to expect. What Eq. \ref{eq:Aronov1} neglects is the possibility of
spin accumulation in the ferromagnet. As we will see later, spin
indeed accumulates in the ferromagnet, strongly modifying the above
estimate for $j_{s0}$. Another simplification we made is to suppose
that the spin is preserved during crossing the interface. This
approximation is actually quite good and will be used in the
standard model as well.

Knowing the spin current at the interface, we can focus on the $N$
region. What happens to the spin which crosses the interface? Unlike
charge, spin is not conserved. Spin relaxes to the equilibrium value
(which is zero in $N$) due to spin-flip scattering and other
spin-randomizing processes. As a result, the motion of the spin in
the presence of spin current will be diffusive.\footnote{In general,
the motion will be a combination of drift and diffusion. At
reasonable electric fields driving the electric current the drift is
much smaller than diffusion and can be neglected.} For the spin
density $s(x)$ in the $N$ region we can then write a diffusion equation
\begin{equation} \label{eq:Aronov2}
\frac{d^2 s}{dx^2} = \frac{s}{L_s^2},
\end{equation}
where $L_s$ is the spin diffusion length in the nonmagnetic conductor. In terms of diffusivity $D$ and
the spin relaxation time $\tau_s$ the spin diffusion length is given as
\begin{equation}
L_s = \sqrt{D \tau_s}.
\end{equation}
The diffusion equation has a general solution,
\begin{equation} \label{eq:Aronov3}
s(x) = s_0 e^{-x/L_s},
\end{equation}
where $s_0= s(0)$ is the spin density at the interface, $x=0$. Above
we applied the physical condition that $s(\infty)=0$.

What remains is to connect the spin density $s_0$ with the spin
current $j_{s0}$. Since the transport of spin is diffusive, the spin
current is
\begin{equation}
j_s = (-e) \times -D \frac{ds}{dx}.
\end{equation}
Note that we define the spin current as the electric current
corresponding to the spin flow---that is why the multiplication by
$-e$ above.  At $x=0$, using Eq. \ref{eq:Aronov3}, we obtain
\begin{equation}
j_s = \frac{-e D}{L_s} s_0 e^{-x/L_s}.
\end{equation}
Assuming that the spin current is continuous across the interface (spin relaxation
is absent there), $j_s(0) = j_{s0}$, we find
\begin{equation}
s_0 = j_{s0}\frac{L_s}{-e D}.
\end{equation}
The full spin density profile in $N$ is given by
\begin{equation}
s(x) = j_{s0}\frac{L_s}{-e D} e^{-x/L_s}.
\end{equation}
The total amount of accumulated spin is
\begin{equation}
s_{\rm acc} = \int_0^\infty s(x) dx = \frac{j_{s_0}L_s^2}{-e D} =
\frac{j_{s0}}{-e} \tau_s.
\end{equation}
In effect, the spin is pumped into the $N$ region. The steady state is achieved by
spin relaxation: The more pumping and the less spin relaxation, the
higher is the spin accumulation.\footnote{Think of inflating a raptured balloon:
the more you blow and the tinier is the hole the bigger the balloon gets. The rapture
symbolizes spin relaxation.}

\section{Spin-polarized transport: concepts and definitions}

\paragraph{Quasichemical potentials.} In thermodynamic equilibrium the chemical
potential $\eta$ throughout the electronic system is uniform, determining the
electron density
\begin{equation} \label{eq:QP1}
n_0(\eta) = \int d\epsilon g(\epsilon) f_0(\epsilon),
\end{equation}
where $g(\epsilon)$ is the electronic density of states at the
energy $\epsilon$ and $f_0$ is the equilibrium Fermi-Dirac
distribution function at a given temperature $T$,
\begin{equation}
f_0(\epsilon) = \frac{1}{\exp(\epsilon - \eta)/k_B T +1}.
\end{equation}
In the presence of an electrostatic potential $\phi(x)$ giving rise
to electric current due to the electric field $E= -\nabla \phi$
inside the conductor, the chemical potential is no longer uniform
(the system is no longer an equilibrium one):
\begin{equation}
\eta \to \eta + e\mu(x),
\end{equation}
where the space dependent addition $\mu(x)$ is the {\it
quasichemical potential}. Since typically the momentum relaxes on
length scales smaller than the variation of $\phi$, we can assume
the local nonequilibrium electron distribution function to be only
energy dependent,
\begin{equation}
f(\epsilon, x) = f_0[\epsilon - e\phi(x) - \eta  -e \mu(x)].
\end{equation}
Then the nonequilibrium electron density is
\begin{equation} \label{eq:QP6}
n(x) = \int d\epsilon g(\epsilon) f(\epsilon, x) = n_0(\eta + e\mu +
e \phi).
\end{equation}

\paragraph{Local charge neutrality.} We make the assumption that charge does not accumulate
inside the conductor under bias $\phi$. This is an excellent
approximation for metals and highly doped (degenerate)
semiconductors. On the other hand, charge can be injected and
accumulated in nondegenerate semiconductors due to the large
screening length. For such cases the standard spin injection model
does not apply. The local charge neutrality means that
\begin{equation}
n(x) = n_0.
\end{equation}
This gives the general condition,
\begin{equation}\label{eq:LCN2}
\mu(x) = -\phi(x).
\end{equation}
The quasichemical potential fully balances the electrostatic
potential.

\paragraph{Electric current.} The electric current comprises the drift current, proportional to
the electric field $E=-\nabla \phi$, and the diffusion current, proportional to the gradient
of the electron density $\nabla n$:
\begin{equation}\label{eq:EC1}
j = \sigma E + e D \nabla n.
\end{equation}
The two proportionality parameters are conductivity $\sigma$ and
diffusivity, $D$. Due to charge neutrality the diffusion current is
absent. We will keep it in the discussion as diffusion will be
present in the spin flow. Using Eq. \ref{eq:QP6}, we write
\begin{equation}
\nabla n = \frac{\partial n_0}{\partial \eta} e\nabla \phi +
\frac{\partial n_0}{\partial \eta} e\nabla \mu.
\end{equation}
Substituting to Eq. \ref{eq:EC1} gives
\begin{equation}
j = \left (-\sigma + e^2 D \frac{\partial n_0}{\partial
\eta} \right ) \nabla \phi + e^2 D \frac{\partial n_0}{\partial \eta} \nabla
\mu.
\end{equation}
There are two important consequences of this equation. First, if
the chemical potential is uniform, $\nabla \mu =0$, the current has
to vanish. This gives the condition on the conductivity,
\begin{equation} \label{eq:Einstein}
{\sigma = e^2 D \frac{\partial n_0}{\partial \eta},}
\end{equation}
known as the Einstein relation. To a good approximation $\partial n_0/\partial \eta
= g(\eta)$, where $g(\eta)$ is the electron density of states at the Fermi level.
Second, using the Einstein relation, the electric current is expressed through the
quasichemical potential only,
\begin{equation} \label{eq:EC6}
{j = \sigma \nabla \mu.}
\end{equation}
This equation generalizes the familiar $j=\sigma E$ to situations
with diffusive currents. The gradient of $\mu$ carries information
on both drift and diffusion.

In a steady state, the continuity of the electric current requires
that
\begin{equation} \label{eq:EC10}
\nabla j = 0,
\end{equation}
that is, the current is uniform. We can also identify the total
increase of the quasichemical potential across the system with
applied voltage. Indeed, for a uniform system of length $L$
integration of Eq. \ref{eq:EC6} gives
\begin{equation}
\Delta \mu = \frac{L}{\sigma} j = {\cal R} j,
\end{equation}
where $\cal{R}$ is the electric resistance of the
system.\footnote{We consider conductors of a unit area cross
section. For rectangular conductors of cross-sectional area $S$ all
the resistances that appear in this article should be divided by
$S$.}

\paragraph{Contact resistance.} At sharp contacts the chemical potential need
not be continuous. Instead of Eq. \ref{eq:EC6} we write
\begin{equation} \label{eq:CR1}
{j = \Sigma \Delta \mu,}
\end{equation}
in which $\Sigma$ is the {\it contact conductance} and $\Delta \mu$
is the increase of the chemical potential across the interface. The
contact electrical resistance is
\begin{equation}
{\cal R}_c = \frac{1}{\Sigma_c}.
\end{equation}

\vspace{0.4cm}
\hspace{0.05\textwidth}
\begin{minipage}{0.9\textwidth}{\small {\bf Problem.} Consider two conductors,
A and B, forming a junction with contact resistance ${\cal R}_c$.
The conductivities of A and B are $\sigma_A$ and $\sigma_B$.
Integrate Eq. \ref{eq:EC6} for each conductor, and apply the
condition of the electric current continuity together with Eq.
\ref{eq:CR1} to obtain $j$ as a function of the applied voltage.
What is the total junction resistance? The standard model of spin
injection goes in the same spirit as this exercise.}
\end{minipage}
\vspace{0.4cm}

\paragraph{Spin density and spin polarization.} Consider a conductor with the electron density $n$.
This density comprises
the densities of spin up and spin down electrons:
\begin{equation}
n= n_{\uparrow} + n_{\downarrow}.
\end{equation}
We define the spin density as
\begin{equation}
s = s_{\uparrow} - s_{\downarrow}.
\end{equation}
A relative difference between the spin up and spin down densities is the
spin polarization of the density,
\begin{equation}
P_n = \frac{s}{n}.
\end{equation}
We add the label $n$ to stress that we speak about the density spin polarization. For
a general spin-resolved quantity $X$, we will have
\begin{equation} \label{eq:SD6}
P_X = \frac{X_\uparrow - X_\downarrow}{X_\uparrow + X_\downarrow},
\end{equation}
and call it the ``spin polarization of X.''

\paragraph{Spin accumulation.} Let us allow for different densities of states $g_\uparrow$
and $g_\downarrow$ at the Fermi level, as well as different
quasichemical potentials $\mu_\uparrow$ and $\mu_\downarrow$ for
spin up and spin down electrons. The equilibrium chemical potential
$\eta$ is the same for both spin species.\footnote{The energy can
flow between spin up and down electrons leading to a common
temperature. Similarly, spin-flip processes lead to exchange of
particles among the two spin pools, giving a unique equilibrium
chemical potential.} Then
\begin{eqnarray}
n_\uparrow(x) &=& n_{\uparrow 0} \left (\eta + e\mu_\uparrow +e\phi
\right ) \approx n_{\uparrow 0} + \frac{\partial n_{\uparrow
0}}{\partial \eta } \left (e\mu_\uparrow + e\phi  \right ), \\
n_\downarrow (x)&=& n_{\downarrow 0} \left (\eta + e\mu_\downarrow
+e\phi \right ) \approx n_{\downarrow 0} + \frac{\partial
n_{\downarrow 0}}{\partial \eta } \left (e\mu_\downarrow + e\phi
\right ),
\end{eqnarray}
where we have expanded the nonequilibrium densities assuming that
$\mu + \phi$ is much smaller than the equilibrium chemical potential
$\eta$; this is a good approximation as it is the electrons close to
the Fermi level that contribute to spin accumulation in degenerate
conductors.  Since $\partial n_0/\partial \eta = g$, we find
\begin{eqnarray} \label{eq:SA3}
n_\uparrow(x) & = & n_{\uparrow 0} + g_\uparrow e\mu_\uparrow +
g_\uparrow e\phi, \\ \label{eq:SA4} n_\downarrow (x) & = &
n_{\downarrow 0} + g_\downarrow e\mu_\downarrow + g_\downarrow
e\phi.
\end{eqnarray}
The local charge neutrality, $n_\uparrow + n_\downarrow = n_0$, then
leads to the condition,
\begin{equation} \label{eq:SA5}
g(\mu + \phi) + g_s \mu_s =0,
\end{equation}
where
\begin{eqnarray}
g & =&  g_\uparrow + g_\downarrow \\
g_s & =&  g_\uparrow - g_\downarrow.
\end{eqnarray}
For nonmagnetic conductors $g_s=0$, recovering Eq. \ref{eq:LCN2}. From Eqs. \ref{eq:SA3}
and \ref{eq:SA4}, using the charge neutrality Eq. \ref{eq:SA5}, we obtain
for the spin density
\begin{eqnarray}
s  =  s_0 + eg_s (\mu+\phi) + eg \mu_s = s_0 + 4  e \mu_s \frac{g_\uparrow g_\downarrow}{g}.
\end{eqnarray}
Here we denoted the quasichemical and {\it spin
quasichemical}\footnote{Beware of factors of ``2''. In the literature
$\mu_s$ is sometimes defined by the plain difference $\mu_\uparrow -
\mu_\downarrow$.} potentials
\begin{eqnarray}
\mu & = & (\mu_\uparrow + \mu_\downarrow)/2, \\
\mu_s & = & (\mu_\uparrow - \mu_\downarrow)/2.
\end{eqnarray}
The accumulated nonequilibrium spin $\delta s$ defined by
\begin{eqnarray}
s = s_0 + \delta s,
\end{eqnarray}
is then
\begin{equation} \label{eq:SA20}
\delta s = 4e\frac{g_\uparrow g_\downarrow}{g} \mu_s.
\end{equation}
Both the nonequilibrium spin density $\delta s$ and the spin
quasichemical potential $\mu_s$ are often termed {\it spin
accumulation}.

\paragraph{Charge and spin currents.} Charge current is the total electric current carried
by spin up and spin down electrons,
\begin{equation}
j = j_\uparrow + j_\downarrow.
\end{equation}
By contrast, spin current is the difference between the electric
currents carried by spin up and spin down electrons:
\begin{equation}
j_s = j_\uparrow - j_\downarrow.
\end{equation}
The two spin components of the electric current are given by
\begin{eqnarray}
j_\uparrow & = & \sigma_\uparrow \nabla \mu_\uparrow, \\
j_\downarrow & = & \sigma_\downarrow \nabla \mu_\downarrow.
\end{eqnarray}
We have labeled the conductivities and the quasichemical potentials with the
corresponding spin index. In nonmagnetic conductors $\sigma_\uparrow = \sigma_\downarrow$.
Let us introduce the charge and spin conductivities
as follow:
\begin{eqnarray}
\sigma & = & \sigma_\uparrow + \sigma_\downarrow, \\
\sigma_s & = &  \sigma_\uparrow - \sigma_\downarrow.
\end{eqnarray}
The electric charge and spin currents become
\begin{eqnarray} \label{eq:SC7}
j & = & \sigma \nabla \mu + \sigma_s \nabla\mu_s, \\ \label{eq:SC8}
j_s & = & \sigma_s \nabla \mu + \sigma \nabla \mu_s.
\end{eqnarray}
For a nonmagnetic conductor $\sigma_s=0$ and the charge and spin
currents decouple; the charge current is driven by the gradient
of the quasichemical potential while the spin current is driven
by the gradient of the spin accumulation. In a ferromagnetic
conductor $\sigma_s \ne 0$ and a gradient in spin accumulation
can cause a charge current. Similarly, a gradient in the quasichemical
potential alone would cause a spin current.

\paragraph{Current spin polarization.} The spin polarization of the
electric current $P_j$ is defined according to Eq. \ref{eq:SD6},
\begin{equation}
P_j = \frac{j_\uparrow - j_\downarrow}{j_\uparrow + j_\downarrow} = \frac{j_s}{j}.
\end{equation}
Extract $\nabla \mu$ from Eq. \ref{eq:SC7},
\begin{equation} \label{eq:CSP2}
\nabla \mu = \frac{1}{\sigma } \left (j - \sigma_s \nabla \mu_s
\right ),
\end{equation}
and substitute into Eq. \ref{eq:SC8}:
\begin{eqnarray} \label{eq:SC9}
j_s  =  \sigma_s \nabla \mu + \sigma \nabla \mu_s = P_\sigma j + 4
\frac{\sigma_\uparrow \sigma_\downarrow}{\sigma} \nabla \mu_s.
\end{eqnarray}
Here $P_\sigma$ is the conductivity spin polarization,
\begin{equation}
P_\sigma = \frac{\sigma_\uparrow -
\sigma_\downarrow}{\sigma_\uparrow + \sigma_\downarrow} =
\frac{\sigma_s}{\sigma}.
\end{equation}
The spin and charge currents are coupled through $P_\sigma$. Finally, the
current spin polarization is
\begin{equation} \label{eq:efficiency}
 {P_j = \frac{j_s}{j} = P_\sigma + \frac{1}{j} 4\nabla \mu_s
 \frac{\sigma_\uparrow \sigma_\downarrow}{\sigma} }.
 \end{equation}
In nonmagnetic conductors $P_\sigma =0$ and spin current is due to
the gradient in spin accumulation only.

\paragraph{Spin-polarized currents in contacts.} The above formalism can
be rewritten for contacts with discrete jumps of the quasichemical potentials. Following
Eq. \ref{eq:CR1}, the spin-resolved currents are
\begin{eqnarray}
j_\uparrow &=& \Sigma_\uparrow \Delta \mu_\uparrow, \\
j_\downarrow &=& \Sigma_\downarrow \Delta \mu_\downarrow.
\end{eqnarray}
Defining the contact charge and spin conductances as
\begin{eqnarray}
\Sigma & = & \Sigma_\uparrow + \Sigma_\downarrow, \\
\Sigma_s & = & \Sigma_\uparrow - \Sigma_\downarrow,
\end{eqnarray}
we can write
\begin{eqnarray}
j & = & \Sigma \Delta \mu + \Sigma_s \Delta \mu_s, \\
j_s &=  & \Sigma_s \Delta \mu + \Sigma \Delta \mu_s.
\end{eqnarray}
Going through similar steps as above of Eq. \ref{eq:efficiency}, we
obtain for the spin current polarization in the contact
\begin{equation}\label{eq:PjC}
{ P_{jc} = P_\Sigma + \frac{1}{j} \frac{\Delta \mu_s(0)}{R_c}.}
\end{equation}
Here
\begin{equation}
P_\Sigma = \frac{\Sigma_\uparrow -
\Sigma_\downarrow}{\Sigma_\uparrow + \Sigma_\downarrow} =
\frac{\Sigma_s}{\Sigma},
\end{equation}
is the contact spin conductance polarization and
\begin{equation} \label{eq:Rc1}
R_c = \frac{\Sigma}{4\Sigma_\uparrow \Sigma_\downarrow},
\end{equation}
is the effective contact resistance,\footnote{This is the first of a
series of effective resistances which appear in the spin injection
problem. To distinguish them from the corresponding electrical
resistances we use calligraphic symbols for the latter.} determining
the drop of the spin accumulation across the contact; $R_c$ is a
quarter of the series resistance of the spin up and spin down
contact resistances. In a spin unpolarized contact $R_c = {\cal R}_c
= 1/\Sigma$.

\paragraph{Diffusion of spin accumulation.} In nonmagnetic systems it is
sufficient to use the continuity of the charge current, Eq.
\ref{eq:EC10}, to find the profile of the quasichemical potential
$\mu(x)$. In the presence of spin polarization, we need a continuity
condition for the spin current as well; the continuity of the charge
current remains unchanged: $\nabla j=0$. Since, unlike charge, spin
is not conserved, the continuity equation for the spin current is
\begin{equation}
\nabla j_s  = e\frac{\delta s }{\tau_s},
\end{equation}
where $\delta s$ is the deviation of the spin density from its
equilibrium value: $s = s_{\rm eq}+ \delta s$.
The divergence of the spin current is proportional to the rate of spin relaxation $1/\tau_s$,
with $\tau_s$ denoting the spin relaxation time. On one hand,
\begin{equation}
\nabla j_s = e\frac{\delta s}{\tau_s} = 4e^2 \mu_s
\frac{g_\uparrow,
g_\downarrow}{g} \frac{1}{\tau_s}
\end{equation}
where we used Eq. \ref{eq:SA20} for $\delta s$. On the other hand,
Eq. \ref{eq:SC9} gives
\begin{equation}
\nabla j_s = \nabla \left ( P_\sigma j + \nabla \mu_s
\frac{4\sigma_\uparrow\sigma_\downarrow}{\sigma} \right ) =
4\frac{\sigma_\uparrow \sigma_\downarrow}{\sigma} \nabla^2 \mu_s.
\end{equation}
Comparing the two we get the following diffusion equation for
spin accumulation:
\begin{equation} \label{eq:DSA10}
{\nabla^2 \mu_s = \frac{\mu_s}{L_s^2},}
\end{equation}
where the generalized spin diffusion length $L_s$ is
\begin{equation}
L_s = \sqrt{\overline{D}\tau_s},
\end{equation}
and the generalized diffusivity
\begin{equation}
\overline{D} = \frac{g}{g_\uparrow/D_\downarrow +
g_\downarrow/D_\uparrow}.
\end{equation}
In a nonmagnetic conductor $\overline{D} = D$. Representative spin
relaxation times $\tau_s$ in nonmagnetic metals and semiconductors are
nanoseconds, and spin diffusion lengths micrometers. In ferromagnetic
conductors these quantities are smaller by several orders of magnitude.

\paragraph{Spin-charge coupling.} Let us write Eq. \ref{eq:CSP2} as
\begin{equation}
\nabla \mu = \frac{j}{\sigma} - P_\sigma \nabla \mu_s,
\end{equation}
and integrate it over a homogeneous region of a conductor:
\begin{equation} \label{eq:SCC9}
\Delta \mu = j {\cal R} - P_\sigma \Delta \mu_s,
\end{equation}
where $\cal R$ is the electrical resistances of the region. Consider
a homogeneous ferromagnetic conductor of length $L \gg L_s$,
stretching from $x=-L$ to $x=0$. Assume that at $x=0$ there is a
spin accumulation $\mu_s(0)$. Applying the above equation gives
\begin{equation}
\mu(0) - \mu(-L) = j {\cal R} - P_\sigma \mu_s (0),
\end{equation}
where the conductor's electric resistance is ${\cal R}=L/\sigma$ and
we assumed absence of spin accumulation at $x = -L$.  In a
nonmagnetic conductor $P_\sigma=0$ and the increase of the
quasichemical potential is due to the charge current flow only. In a
ferromagnetic conductor the increase is also due to the spin
accumulation. In an open circuit ($j=0$) the increase in the
quasichemical potential is
\begin{equation}
\mu(0) - \mu(-L) = - P_\sigma \mu_s (0),
\end{equation}
This increase defines the electromotive force (emf) per unit
charge\footnote{The spin accumulation at first generates spin
diffusion and the connected electron flow---since we are dealing
with a ferromagnet. In the open circuit a balancing electric field
develops preventing unlimited buildup of charges at the two ends of
the conductor. The resulting emf is the work done by the source of
the spin accumulation in bringing the electrons through the
conductor against the built-up electric field.} generated by the
spin accumulation in the ferromagnetic conductor. Similarly, we can
calculate the corresponding drop in the electric potential,
\begin{equation} \label{eq:SCC6}
\phi(-L) - \phi(0)  = (P_g -  P_\sigma) \mu_s(0),
\end{equation}
where we used the local neutrality condition, Eq. \ref{eq:SA5}. The
density of states spin polarization is
\begin{equation}
P_g = \frac{g_\uparrow - g_\downarrow}{g_\uparrow + g_\downarrow} =
\frac{g_s}{g}.
\end{equation}
Equation \ref{eq:SCC6} is an example of spin-charge coupling: The
presence of a spin accumulation in a conductor with an equilibrium
spin polarization, a nonequilibrium voltage drop (electromotive
force) develops. Electrostatic detection of the voltage drop then
allows to extract the magnitude of the spin accumulation.

\section{The standard model of spin injection:
$F$/$N$ junction}

We pose the following question:

\vspace{0.2cm} {\it Knowing the equilibrium materials parameters of
a ferromagnet ($F$), a nonmagnetic conductor ($N$), as well as the
properties of the contact ($C$) between them, what is the spin
current polarization and spin accumulation in $N$, in the presence
of electric current $j$? } \vspace{0.2cm}

The scheme of the $F$/$N$ junction we consider is in Fig. 1.
The spin current polarization at the contact
is termed {\it spin injection efficiency}. We denote it as $P_j$. To
obtain $P_j$ we need to consider spin-polarized transport separately
in the three regions: $F$, $C$, and $N$. The solutions for the
transport equations will then be connected by suitable continuity
conditions. We also add labels $F$, $C$, and $N$ to the quantities
pertaining to the three regions.

\begin{center}
\begin{figure} \label{fig:FNjunction}
\epsfig{file=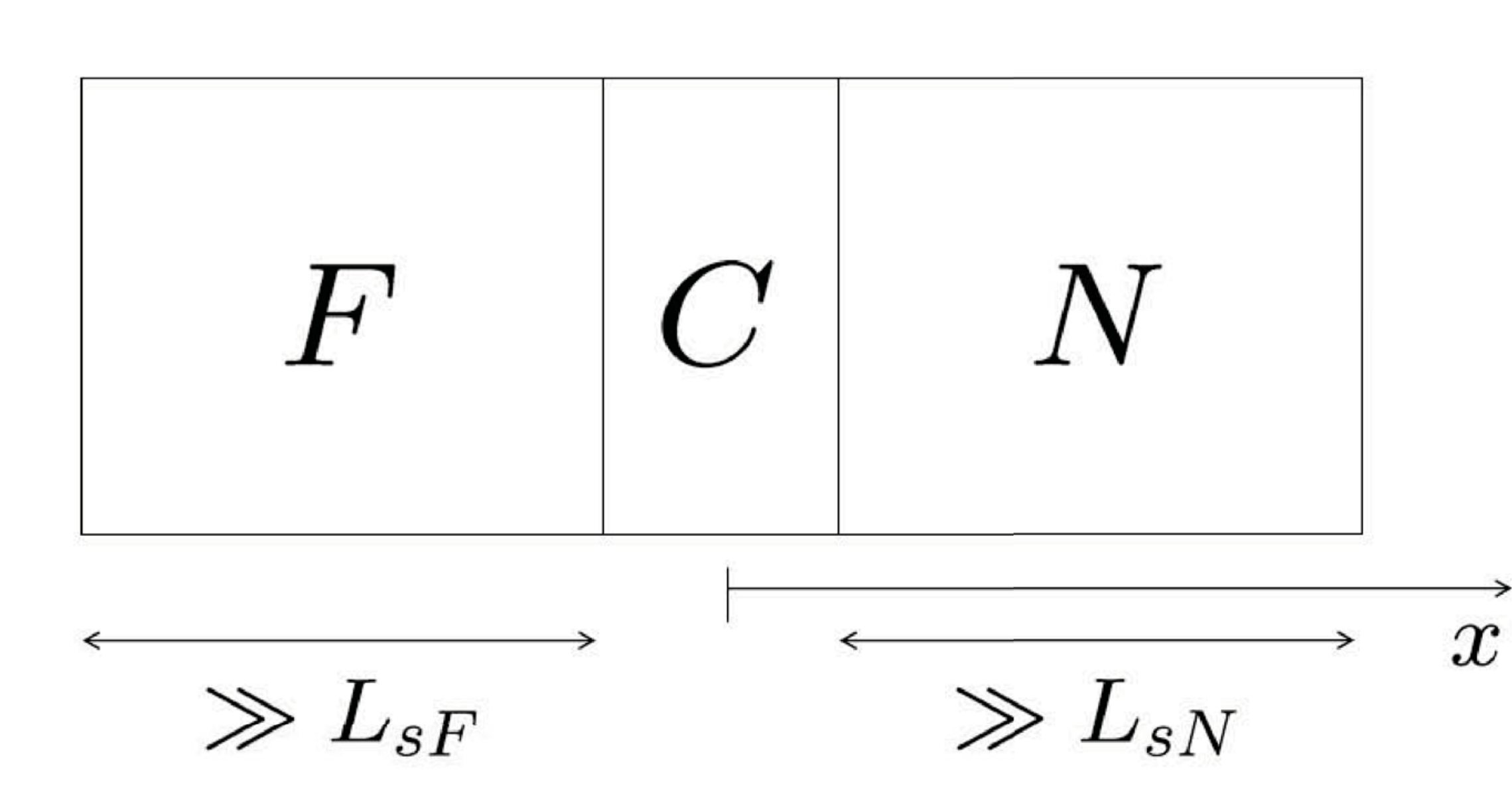,width=0.5\textwidth} \caption{The
$F$/$N$ junction above comprises a ferromagnetic conductor $F$, a
nonmagnetic conductor $N$, as well as the contact $C$ between them
at $x=0$. It is assumed that the widths of the $F$ and $N$ regions
are much larger (we call their size ``$\infty$'') than the
corresponding spin diffusion length. }
\end{figure}
\end{center}

\paragraph{Ferromagnetic conductor.} The ferromagnetic conductor occupies the region $(-\infty, 0)$. The spin
accumulation profile is given by the solution of the diffusion
equation, Eq. \ref{eq:DSA10}, as
\begin{equation}
\mu_{sF} = \mu_{sF}(0) e^{x/L_{sF}}.
\end{equation}
We have applied the condition that there is no spin accumulation at
$x=-\infty$: $\mu_{sF}(-\infty)=0$. This condition is well satisfied
if the length of the ferromagnet, indicated by ``$\infty$'', is much
larger than the spin diffusion length $L_{sF}$. From the above we
have
\begin{equation}
\nabla \mu_{sF}(0) = \frac{\mu_{sF}(0)}{L_{sF}}.
\end{equation}
Substituting to Eq. \ref{eq:efficiency} we obtain the spin current
polarization in the $F$ region of the contact
\begin{equation} \label{eq:FC8}
P_{jF}(0) = P_{\sigma F} + \frac{1}{j} \frac{\mu_{sF}(0)}{R_F},
\end{equation}
where we denote
\begin{equation}
R_F = \frac{\sigma_F}{4\sigma_{F \uparrow}\sigma_{F\downarrow}}
L_{sF},
\end{equation}
the effective resistance of the ferromagnet; $R_F$ is a quarter of
the serial resistance of the spin up and spin down resistances of a
piece of the ferromagnet of length $L_{sF}$. We stress that $R_F$ is
{\it not} the actual resistance of the $F$ region ${\cal R}_F$,
which is
\begin{equation}
{\cal R}_F = \frac{``\infty "}{\sigma_{\uparrow} +
\sigma_{\downarrow}},
\end{equation}
given as a parallel resistance of the two spin channels over the
entire size $``\infty"$ of the ferromagnet. The two resistances,
$R_F$ and ${\cal R}_F$ can be very different!

\paragraph{Nonmagnetic conductor.} In a nonmagnetic conductor the transport and materials
parameters are spin independent and all the equilibrium polarizations, such as $P_\sigma$ or $P_g$, vanish.
The profile of the spin accumulation is
\begin{equation}
\mu_{sN} = \mu_{sN}(0) e^{-x/L_{sF}},
\end{equation}
satisfying the boundary condition $\mu_{sN}(\infty) = 0$. We then have
\begin{equation}
\nabla \mu_{sN}(0) = - \frac{\mu_{sN}(0)}{L_{sN}},
\end{equation}
and the spin current polarization at the contact
\begin{equation} \label{eq:NC8}
P_{jN}(0) = - \frac{1}{j} \frac{\mu_{sN}(0)}{R_N},
\end{equation}
where
\begin{equation}
R_N = \frac{L_{sN}}{\sigma_{N}},
\end{equation}
is the effective resistance of the $N$ region; $R_N$ is the
resistance of a piece of a conductor of size $L_{sN}$. Again,
$R_N$ can be very different from the actual electric resistance of
the $N$ region, ${\cal R}_N$.

\paragraph{Contact region.} The contact region is described by Eq. \ref{eq:PjC}. For
our $F$/$N$ contact the spin current polarization is
\begin{equation} \label{eq:C9}
P_{jc} = P_\Sigma + \frac{1}{j}\frac{\mu_{sN}(0) -
\mu_{sF}(0)}{R_c}.
\end{equation}

\paragraph{Spin injection efficiency.} We have three equations
for the spin current polarizations, Eqs. \ref{eq:FC8}, \ref{eq:NC8},
\ref{eq:C9}, in three different regions. We assume that spin is
conserved across the contact. As a consequence, the spin current
(and thus spin current polarization) is continuous there:
\begin{equation}
P_j \equiv P_{jF}(0) = P_{jN}(0) = P_{jc}.
\end{equation}
Solving this straightforward algebraic problem leads to the
important expression for the spin injection efficiency:
\begin{equation} \label{eq:SIE4}
{ P_j = \frac{R_F P_{\sigma F} + R_c P_\Sigma}{R_F + R_c + R_N } =
\langle P_\sigma \rangle_R.}
\end{equation}
This equation is one of the main results of the standard model of
spin injection. The spin injection efficiency is the weighted
average of the equilibrium spin conductance polarizations of the
system; the weight is the relative effective resistance.

\begin{figure}\label{fig:FNprofile}
\epsfig{file=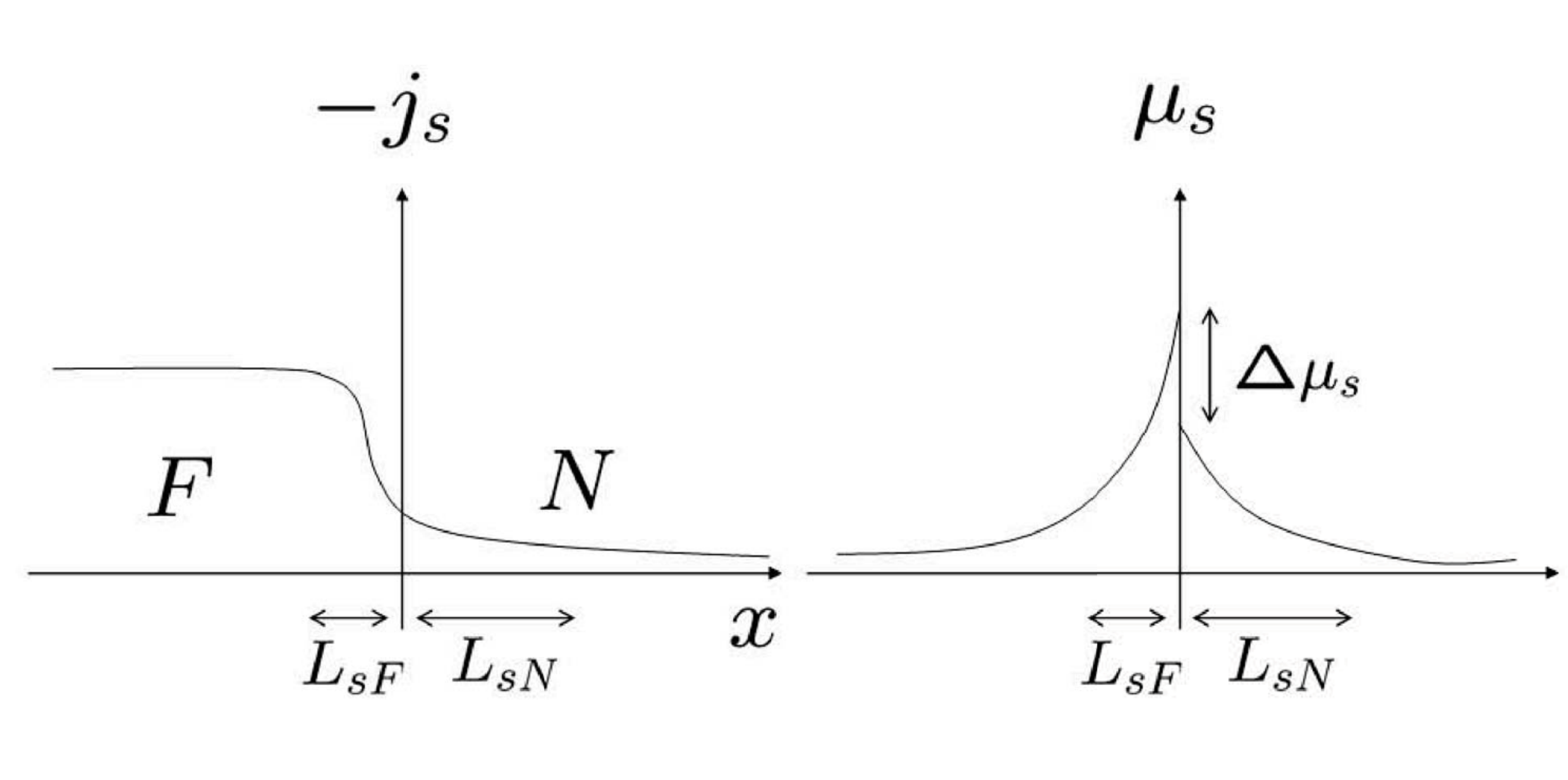,width=0.5\textwidth} \caption{Sketch of the
spatial profile of the spin current $j_s$ and the spin
quasichemical potential $\mu_s$ in an $F$/$N$ junction in the spin
injection regime. While the spin current is continuous throughout
the junction, the spin quasichemical potential experiences a jump at
the contact.}
\end{figure}

\vspace{0.4cm}

\hspace{0.05\textwidth}
\begin{minipage}{0.9\textwidth}{\small {\bf Problem.} {\it $F$/$F$ junction.}
Calculate the spin injection efficiency for a $F$/$F$ junction of
two different ferromagnets. Show that $P_j= \langle P_\sigma
\rangle_R$ still holds. }
\end{minipage}
\vspace{0.4cm}

\paragraph{Spin injection and spin extraction.} Knowing $P_j$ we can calculate
the spin accumulation in the $N$ region,
\begin{equation}
\mu_{sN}(0) = - j P_j R_N,
\end{equation}
and the corresponding spin density polarization,
\begin{equation}
P_n(0) = \frac{s(0)}{n}= e \mu_{sN}(0) \frac{g_N}{n} = - j e R_N
\frac{g_N}{n} P_j.
\end{equation}
Since the spin polarization is proportional to  the electric
current, the electric spin injection is a realization of spin
pumping. In a typical spin injection experiment electrons flow from $F$ to $N$,
so that $j<0$. In this case $P_n(0)$ has the same size as $P_j$ and
we speak of {\it spin injection}. If the electric current is
reversed, $j<0$, electrons from $N$ flow into $F$. Now $P_n(0)$ has
the opposite sign to $P_j$ and we speak of {\it spin extraction}.
For a positive $P_j$, for example, more spin up than spin down
electrons are transported through the contact, leaving a negative
spin density in the $N$ region. A sketch of the profile of the spin
current and spin quasichemical potential across an $F$/$N$ junction
is shown in Fig. 2.

\paragraph{Equivalent circuit.} The standard model of spin injection
can be formulated by a simple equivalent circuit model, shown in
Fig. 3. The model is a parallel circuit
with spin up and spin down channels. Each region is characterized by
the corresponding effective resistance.

\begin{figure} \label{fig:equivalent_circuit}
\epsfig{file=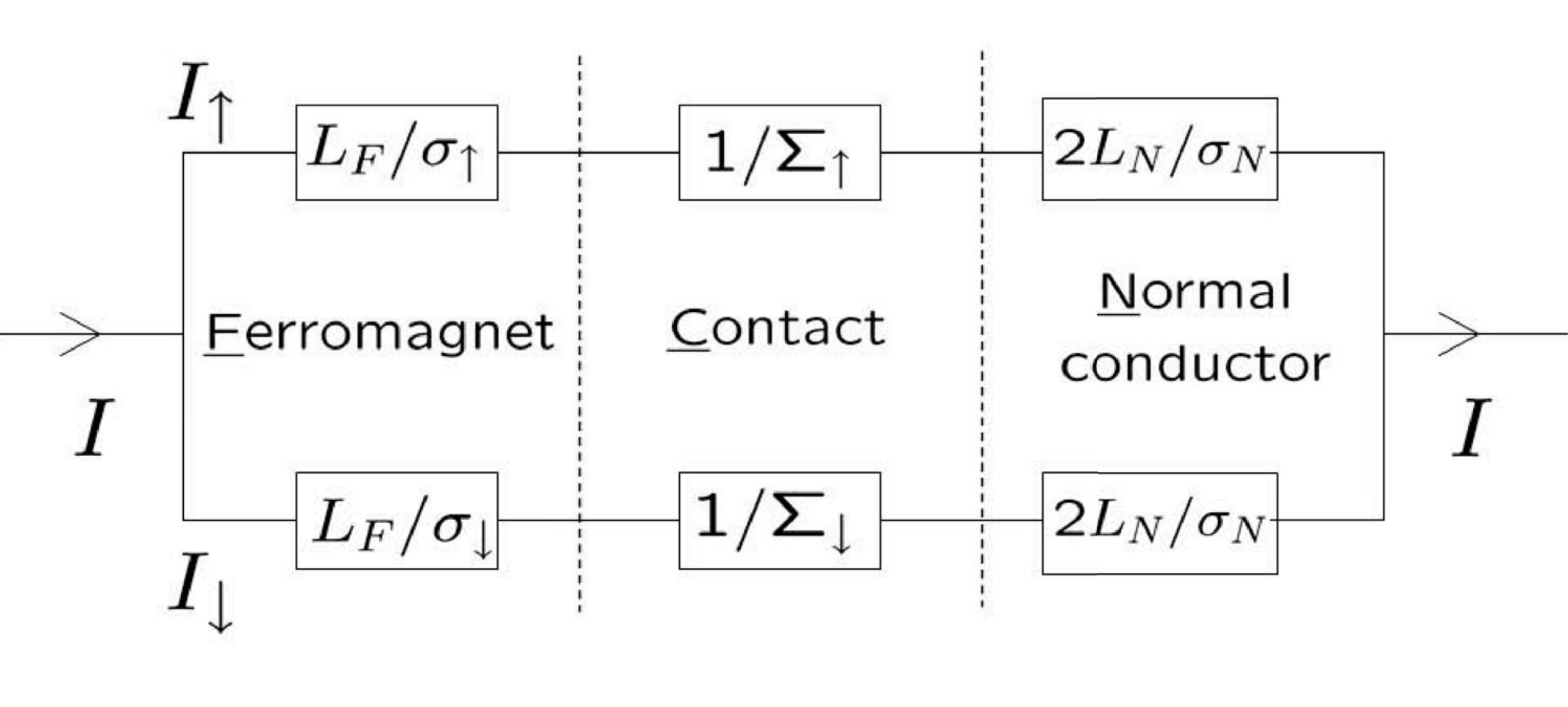,width=0.5\textwidth}
\caption{Equivalent circuit of a $F$/$N$ junction in the spin
injection regime. The electric current splits into spin up and spin
down channels, each with a series of three effective resistances as
indicated.}
\end{figure}

\vspace{0.4cm}
\hspace{0.05\textwidth}
\begin{minipage}{0.9\textwidth}{\small {\bf Problem.} Calculate $I_s = I_\uparrow - I_\downarrow$ from the
equivalent circuit model and show that $P_j = I_s/I$ agrees with Eq.
\ref{eq:SIE4}.}
\end{minipage}

\vspace{0.4cm} \hspace{0.05\textwidth}
\begin{minipage}{0.9\textwidth}{\small {\bf Problem.} Formulate the equivalent circuit model for a F/F
junction.}
\end{minipage}

\vspace{0.4cm} \hspace{0.05\textwidth}
\begin{minipage}{0.9\textwidth}{\small {\bf Problem.} {\it $F$/$N$/$N$ junction}.
Consider electrical spin injection in an $F$/$N$/$N$ junction in
which the two $N$ regions are different (say, GaAs and Si).
Calculate the spin injection efficiency at the $N$/$N$ interface.
What is the spin accumulation at both sides of this interface?
Sketch the profile of the spin accumulation across this junction.}
\end{minipage}
\vspace{0.4cm}

\section{Nonequilibrium resistance and spin bottleneck}

In the absence of spin accumulation the resistance of the F/N
junction is ${\cal R}_F + {\cal R}_N + {\cal R}_c$.
Spin accumulation leads to an additional positive resistance
$\delta {\cal R}$ so that the increase of the quasichemical
potential (which generates the emf) is
\begin{equation}
\mu_N(\infty) - \mu_F(-\infty) = ({\cal R}_F + {\cal R}_N + {\cal
R}_c + \delta {\cal R})j.
\end{equation}
Let us apply Eq. \ref{eq:SCC9} to the three regions, $F$, $C$, and
$N$, successively:
\begin{eqnarray}
\mu_F(0) - \mu_F(-\infty) & = & j {\cal R}_F - P_{\sigma F}
\mu_{sF}(0),
\\
\mu_N(0) - \mu_F(0) & = & j {\cal R}_c - P_\Sigma \left [\mu_{sN}(0) - \mu_{sF}(0)  \right ], \\
\mu_N(\infty) - \mu_N(0) & = & j {\cal R}_N.
\end{eqnarray}
We have used that $\mu_{sF} (-\infty) = 0$. Summing up the above
equations gives for the nonequilibrium resistance
\begin{equation}
\delta {\cal R} = - ( P_{\sigma F} - P_\Sigma) \mu_{sF}(0) -
P_\Sigma \mu_{sN}(0).
\end{equation}
Expressing the spin quasichemical potentials at $x=0$ in terms of
the spin injection efficiency, see Eqs. \ref{eq:FC8} and
\ref{eq:NC8},
\begin{eqnarray}
\mu_{sF} (0) & = & j R_F (P_{j} - P_{\sigma F}), \\
\mu_{sN}(0) & = & - j R_N P_j,
\end{eqnarray}
we get
\begin{equation}
\delta {\cal R} = - P_\Sigma (P_j - P_\Sigma) R_c - P_{\sigma F}
(P_j - P_{\sigma F})R_F.
\end{equation}
Using the expression for $P_j$ in Eq. \ref{eq:SIE4}, we obtain the
final result
\begin{equation}
\delta {\cal R} = \frac{R_N(P_\Sigma^2 R_c + P_{\sigma F}^2 R_F) +
R_F R_c (P_{\sigma F} - P_\Sigma)^2}{R_F + R_c + R_N} > 0.
\end{equation}
The nonequilibrium resistance is always positive!

\vspace{0.4cm} \hspace{0.05\textwidth}
\begin{minipage}{0.9\textwidth}{\small {\bf Problem.} Obtain the nonequilibrium
resistance $\delta {\cal R}$ from the equivalent circuit model in
Fig. 3, as $\delta {\cal R} = {\cal R} -
L_{sF}/\sigma_F - L_{sN}/\sigma_N$.}
\end{minipage}
\vspace{0.4cm}

What is the reason behind the additional positive resistance due to
spin accumulation? As the nonequilibrium spin piles up in the
ferromagnet and the spin-polarizing contact region, the spin
diffusion there pushes the electrons against the flow of the
electric current. Indeed, the electric current brings electrons from
the spin-polarized region to the nonmagnetic conductor, while the
spin diffusion in the ferromagnet and the contact drives them back
to the ferromagnet. This {\it spin bottleneck effect} causes the
additional electrical resistance of the junction.

\section{Transparent and tunnel contacts, conductivity mismatch}

Two important cases are analyzed: transparent and tunnel contacts.

\paragraph{Transparent contacts.} By transparent contacts we mean
the condition
\begin{equation}
R_c \ll R_N, R_F.
\end{equation}
This is the case of usual ohmic contacts between two metals or
degenerate semiconductors. Using our results for the $F$/$N$
junction, a transparent contact is characterized by the spin
efficiency
\begin{equation}
P_j  =  \frac{R_F}{R_F + R_N} P_{\sigma F}.
\end{equation}
For metals $\sigma_F$ is usually somewhat less than $ \sigma_N$, as
$L_{sN} \gg L_{sF}$. We then get
\begin{equation}
P_j \approx (\sigma_N/\sigma_F) (L_{sF}/L_{sN}).
\end{equation}
If  $N$ is a semiconductor while $F$ is a metal, so that $\sigma_N
\ll \sigma_F$, the spin injection efficiency is greatly reduced.
This inefficiency of the spin injection from a ferromagnetic metal
to a nonmagnetic semiconductor via a transparent contact is known as
the {\it conductivity mismatch} problem, since it comes from the
greatly different conductivities of the two regions of the
junctions.

The nonequilibrium resistance of a transparent contact is
\begin{equation}
\delta {\cal R} = \frac{R_N R_F}{R_N + R_F} P_{\sigma F}^2.
\end{equation}
Again, since typically $R_N$ is greater than $R_F$,
\begin{equation}
\delta {\cal R} \approx R_F P_{\sigma F}^2 = \frac{L_{sF}}{\sigma_F}
P_{\sigma F}^2.
\end{equation}
In the extreme limit of the conductivity mismatch, the
nonequilibrium resistance will be negligible as compared to the
usual electrical junction resistance which will be dominated by
${\cal R}_N$.

\paragraph{Tunnel contacts.} By tunnel contacts we mean
\begin{equation}
R_c \gg R_{N}, R_{F}.
\end{equation}
The contact dominates the electric properties of the junction. The
spin injection efficiency for a tunnel contact is
\begin{equation}
P_j \approx P_\Sigma.
\end{equation}
The contact also dominates the spin injection efficiency. The
conductance mismatch in tunnel contacts plays no role and spin
injection from a ferromagnetic metal to a nonmagnetic semiconductor
can be highly efficient.

The nonequilibrium resistance of a tunnel contact is
\begin{equation}
\delta {\cal R} = R_N P_\Sigma^2 + R_F (P_{\sigma F} - P_\Sigma)^2.
\end{equation}
This is in general much less than the electric resistance of the
contact, ${\cal R}_c$.

\vspace{0.4cm} \hspace{0.05\textwidth}
\begin{minipage}{0.9\textwidth}{\small {\bf Problem.} {\it Spin accumulation in transparent and
tunnel junctions}. Calculate the spin accumulation $\mu_{sN}(0)$ and
the spin density polarization $P_{\sigma N}(0)$ in a transparent and
a tunnel $F$/$N$ junction. What is the spin density polarization in
the conductivity mismatch problem? Can it be significant?}
\end{minipage}
\vspace{0.4cm}

\hspace{0.05\textwidth}
\begin{minipage}{0.9\textwidth}{\small {\bf Problem.} {\it Equivalent circuit
of the conductivity mismatch problem}. Draw the equivalent circuit
for the conductivity mismatch problem of a transparent $F$/$N$
junction and use it to explain the spin injection inefficiency.}
\end{minipage}
\vspace{0.4cm}

\section{Silsbee-Johnson spin-charge coupling}

Driving electric current through a $F$/$N$ interface generates spin
accumulation by the process of spin injection. The Silsbee-Johnson
spin charge coupling is the inverse of spin injection: emf develops
by the presence of a spin accumulation in the proximity of a
ferromagnetic conductor. We will analyze the coupling in an open
$F$/$N$ junction, that is in the absence of electric current
($j=0$), under the condition of $\mu_{sN}(\infty) \ne 0$ which
models a source of nonequilibrium spin far in the nonmagnetic
region. The scheme is shown in Fig. 4.

\begin{figure} \label{fig:Silsbee-Johnson1}
\epsfig{file=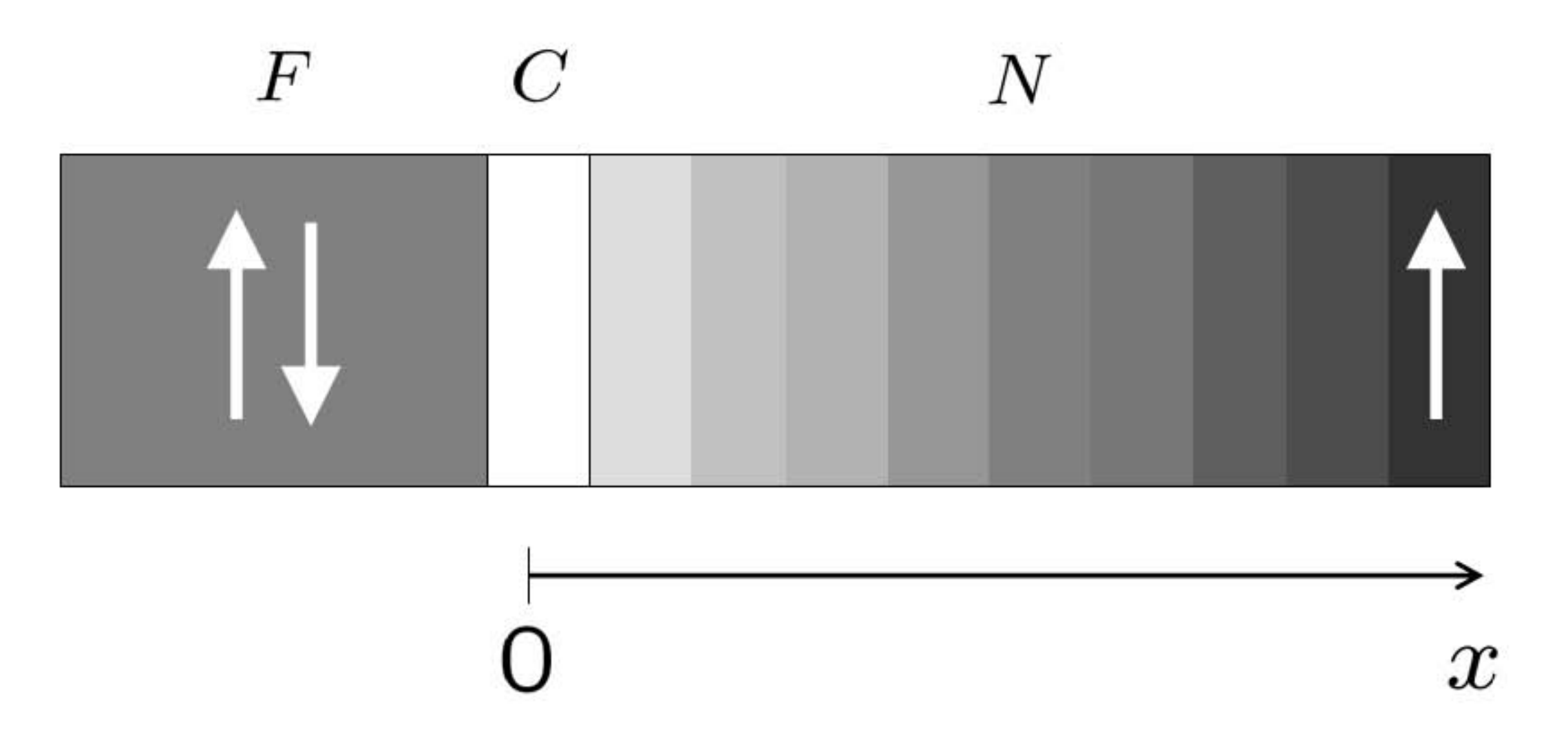,width=0.5\textwidth}
\caption{Scheme of the Silsbee-Johnson spin-charge coupling. A spin
source at the far right of the $N$ region drives spin by diffusion
towards the spin-polarizing contact and ferromagnet. The proximity
of the nonequilibrium spin and the equilibrium spin polarization
gives rise to an emf in the open circuit (or electric current when
the circuit is closed).}
\end{figure}

The induced emf is the increase of the quasichemical potential across the
junction,
\begin{equation}
{\rm emf} = \mu_N(\infty) - \mu_F(-\infty).
\end{equation}
The charge neutrality and the physical condition that $\mu_{sF}(-\infty)=0$ guarantee
that the emf can be detected as a drop of the electric voltage:
\begin{equation}
{\rm emf} = \mu_N(\infty) - \mu_F(-\infty) = \phi_F(-\infty) - \phi_N(\infty).
\end{equation}
Our strategy is to first express the quasichemical potential increase
in terms of the spin accumulations at the contact, and then use the
spin current continuity at the contact as well as the diffusion of the
spin accumulation to find the spin accumulations.

In the absence of electric current we can apply Eq. \ref{eq:SCC9} to
$F$, $C$, and $N$ regions sequentially:
\begin{eqnarray}
\mu_F(0) - \mu_F(-\infty) & = & -P_{\sigma_F} \mu_{sF}(0), \\
\mu_N(0) - \mu_F(0) & = & -P_\Sigma \left [\mu_{sN}(0) - \mu_{sF}(0)\right ], \\
\mu_N(\infty) - \mu_N(0) & = & 0.
\end{eqnarray}
In the nonmagnetic conductor there is no voltage drop associated with
the presence of spin accumulation if $j=0$.
Summing up the above equations gives
\begin{equation}
{\rm emf} = \mu_N(\infty) - \mu_F(-\infty) = \mu_{sF}(0) (P_\Sigma -
P_{\sigma F}) -  \mu_{sN}(0) P_\Sigma.
\end{equation}

\begin{figure} \label{fig:FNscc}
\epsfig{file=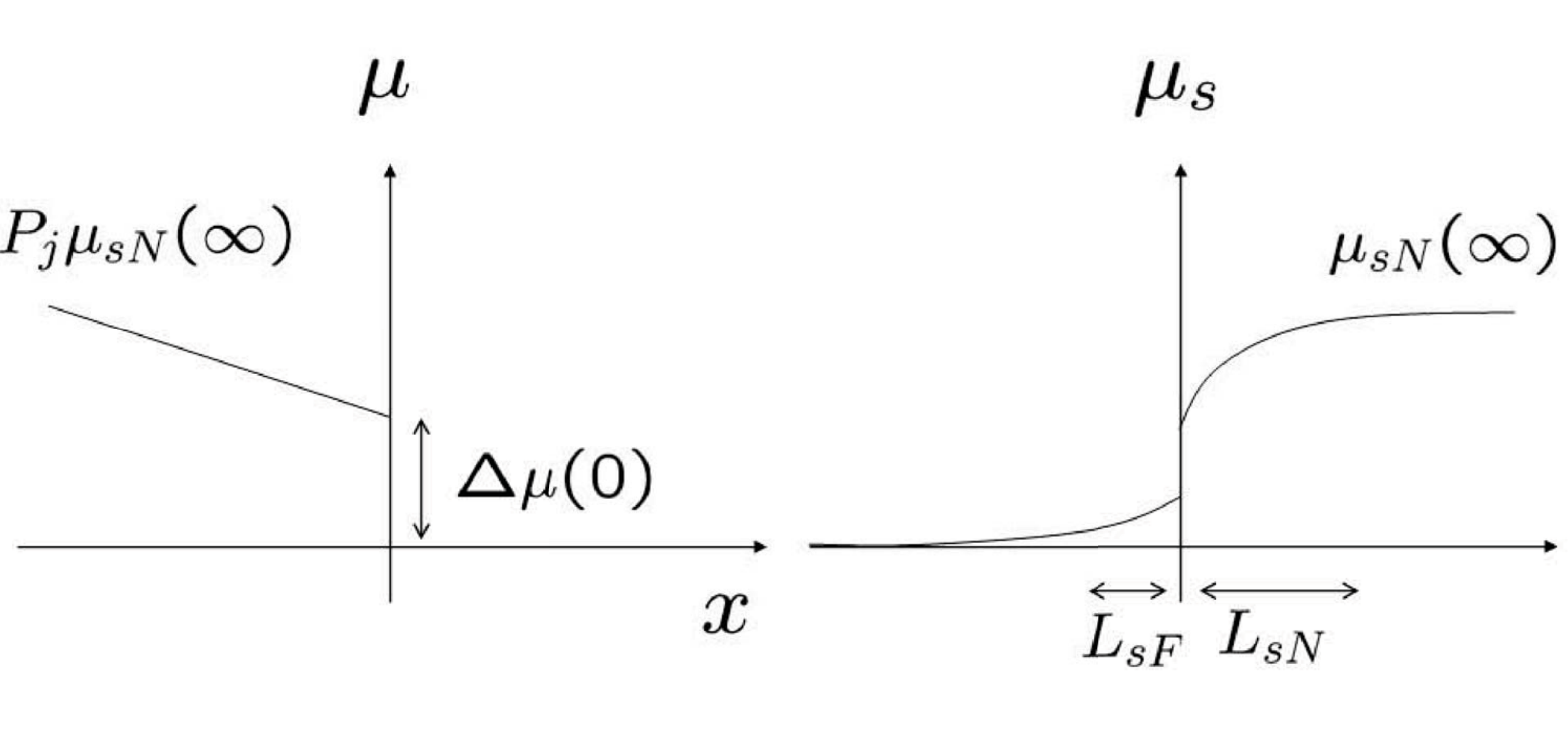,width=0.5\textwidth} \caption{Sketch of the
spatial profile of the quasichemical potential $\mu$ (left) and the
spin quasichemical potential $\mu_s$ (right) in a $F$/$N$ junction
in the spin-charge coupling regime. }
\end{figure}

In the $N$ region, due to the presence of spin accumulation at the
far right, the spin accumulation diffusion profile is
\begin{equation}
\mu_{sN}(x) = \mu_{sN} (\infty) + \left [\mu_{sN}(0) -
\mu_{sN}(\infty)  \right ] e^{-x/L_{sN}},
\end{equation}
as can be verified by direct substitution to the diffusion equation,
Eq. \ref{eq:DSA10}. To calculate the spin current at $x=0$ in the
$N$ region we need the gradient,
\begin{equation}
\nabla \mu_{sN}(0) = -\frac{1}{L_{sN}} \left [\mu_{sN}(0) - \mu_{sN}(\infty)  \right ].
\end{equation}
We are now ready to calculate the spin currents at the interface,
for the three regions. Equation \ref{eq:SC9} gives
\begin{eqnarray} \label{eq:jsN2}
j_{sN}(0) &  = & -\frac{1}{R_N} \left [\mu_{sN}(0) -
\mu_{sN}(\infty) \right ], \\ \label{eq:jsF2} j_{sF}(0) & = &
\frac{1}{R_F} \mu_{sF}(0), \\ \label{eq:jsc2}j_{sc} & = &
\frac{1}{R_c} \left [ \mu_{sN}(0) - \mu_{sF}(0) \right ].
\end{eqnarray}
Assuming that the three spin currents are equal,
\begin{equation}
j_s \equiv j_{sF}(0) = j_{sc} = j_{sN}(0),
\end{equation}
we obtain
\begin{equation}
j_s = \frac{\mu_{sN}\,(\infty)}{R_F + R_c + R_N}.
\end{equation}
The emf is then found from
\begin{equation}
{\rm emf} = (P_{\sigma F} R_F + P_\Sigma R_c) j_s,
\end{equation}
which gives the spin-charge coupling in the final form
\begin{equation}
{\rm emf} = P_j \mu_{sN}(\infty),
\end{equation}
where $P_j$ is the spin injection efficiency of the junction, given
in Eq. \ref{eq:SIE4}. The spin-charge coupling allows electrical
detection of spin accumulation. Placing a ferromagnetic electrode
over a conductor with nonequilibrium spin generates emf in the open
circuit, or electric current if the junction is part of a closed
circuit. The spin accumulation can be generated electrically (see
the section on the nonlocal geometry) or by other means (optically
or by spin resonance). Figure 5 shows the profile of
the quasichemical potentials across the junction.

The origin of the spin-charge coupling can be traced to the presence
of the spin current in the ferromagnet. If $P_\sigma \ne 0$ the spin
current would also induce electric current. In an open circuit there
is instead a balancing emf induced.

\vspace{0.4cm}
\hspace{0.05\textwidth}
\begin{minipage}{0.9\textwidth}{\small {\bf Problem.} Sketch the spatial profiles
of $\mu$, $\mu_s$, and $j_s$ in the $F$/$N$ junction in the Silsbee-Johnson
spin-charge coupling regime.
}
\end{minipage}
\vspace{0.4cm}

\section{Spin injection in $F$/$N$/$F$ junctions}

The same technique which we applied to study the spin injection in
$F$/$N$ junctions is applicable to more general structures. We will
use it to analyze the spin injection in $F$/$N$/$F$ junctions. By
independent switching of the orientations of the magnetizations of
the two $F$ regions, the junctions can be in the parallel ($\uparrow
\uparrow$) or antiparallel ($\uparrow \downarrow$) configurations.
We will in particular be interested in the difference of the
junction electrical resistance for antiparallel and parallel
configurations,
\begin{equation}
\Delta {\cal R} = \delta {\cal R}^{\uparrow \downarrow} - \delta
{\cal R}^{\uparrow \uparrow}.
\end{equation}
This difference contains only contributions of the respective
nonequilibrium resistances.

\begin{figure} \label{fig:FNF}
\epsfig{file=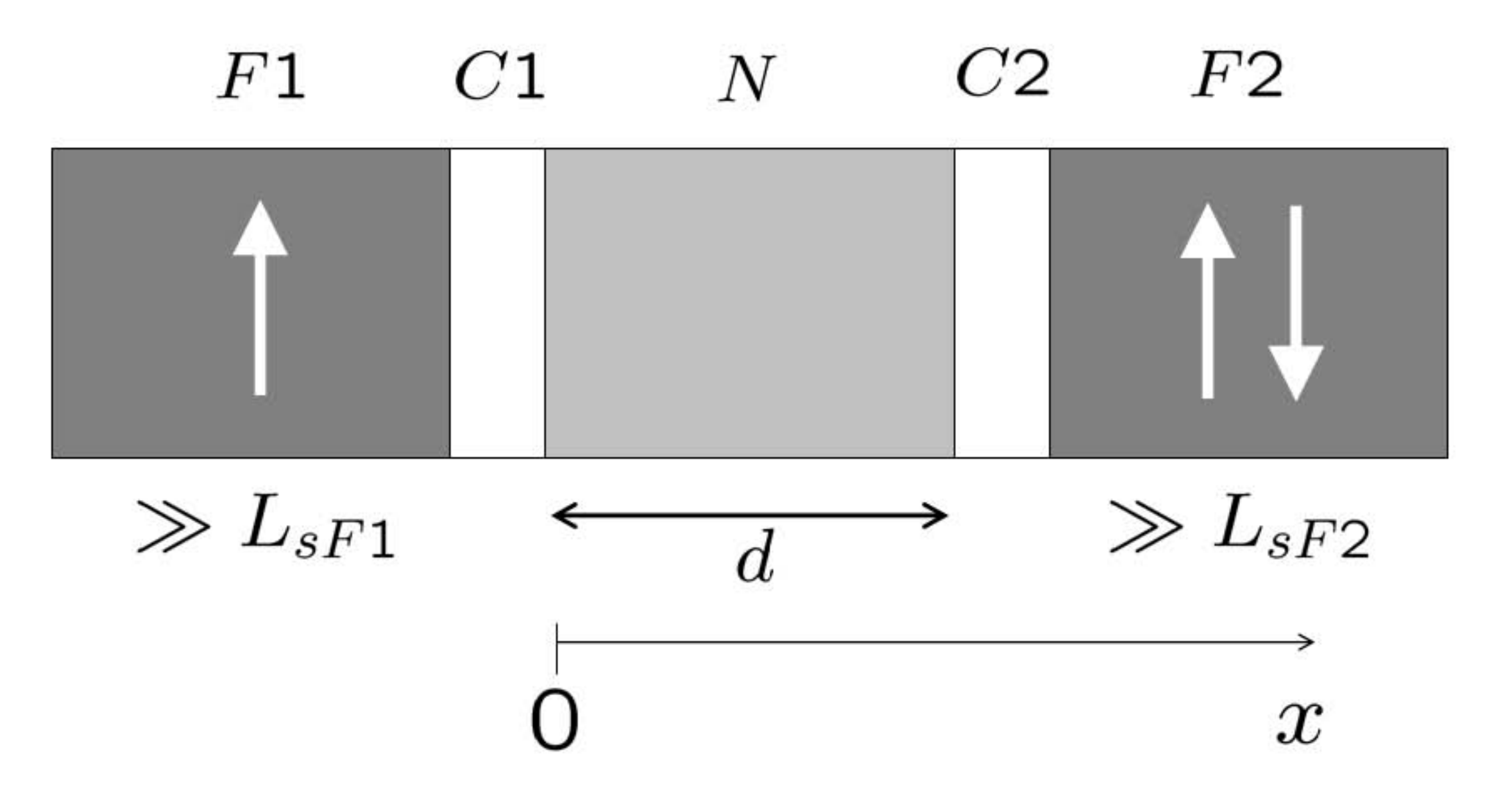,width=0.5\textwidth} \caption{Sketch of an
$F$/$N$/$F$ junction. The magnetization of the $F2$ conductor can be
up or down, giving parallel and antiparallel configurations. }
\end{figure}

\paragraph{Spin injection efficiencies.} The described geometry is
shown in Fig. 6. The width of the nonmagnetic conductor
is $d$. The two ferromagnetic layers are labeled $F1$ and $F2$. We
start with a generic asymmetric configuration in which $F1$ and $F2$
are different. Going through similar steps as in the $F$/$N$
junction, we find that the spin current polarizations in the
ferromagnets at $x=0$ and $x=d$ are
\begin{eqnarray}
P_{jF1}(0) & = & P_{\sigma F1} + \frac{1}{j} \frac{\mu_{sF1}(0)}{R_{F1}}, \\
P_{jF2}(d)  & = & P_{\sigma F2} - \frac{1}{j}
\frac{\mu_{sF2}(d)}{R_{F2}}.
\end{eqnarray}
Similarly, at the two contacts we have
\begin{eqnarray}
P_{jc1} &= &P_{\Sigma 1} + \frac{1}{j}\frac{\Delta \mu_{s}(0)}{R_{c1}}, \\
P_{jc2} &= &P_{\Sigma 2} + \frac{1}{j}\frac{\Delta
\mu_{s}(d)}{R_{c2}}.
\end{eqnarray}
In contrast to the $F$/$N$ junction, the $N$ region is of finite
width $d$. Considering the quasichemical potentials $\mu_{sN}(0)$
and $\mu_{sN}(d)$ as yet unknown boundary conditions, the solution
to the diffusion equation \ref{eq:DSA10} is
\begin{equation} \label{eq:FNF9}
\mu_{sN} (x) = \mu_{sN}(d) \frac{\sinh(x/L_{sN})}{\sinh(d/L_{sN})} -
\mu_{sN}(0) \frac{\sinh\left [(x-d)/L_{sN}\right
]}{\sinh(d/L_{sN})}.
\end{equation}
By evaluating $\nabla\mu_{sN}(0)$ and $\nabla \mu_{sN}(d)$ from the
above equation, we obtain for the spin current polarizations in the
nonmagnetic region,
\begin{eqnarray}
P_{jN}(0) & = &\frac{1}{j R_N} \frac{1}{\sinh(d/L_{sN})} \left [\mu_{sN}(d) - \mu_{sN}(0) \cosh(d/L_{sN})  \right ], \\
P_{jN}(d) & = & \frac{1}{j R_N} \frac{1}{\sinh(d/L_{sN})} \left
[\mu_{sN}(d) \cosh(d/L_{sN}) - \mu_{sN}(0)  \right ].
\end{eqnarray}
The above equations for the spin current polarizations need to be
supplemented by the continuity conditions for the spin currents at
the two contacts:
\begin{eqnarray}
P_{j1} \equiv P_{jF1}(0) = P_{jc1} = P_{jN}(0), \\
P_{j2} \equiv P_{jF2}(d) = P_{jc2} = P_{jN}(d).
\end{eqnarray}
The algebraic system is now complete and we can solve it to obtain the
spin injection efficiencies $P_{j1}$ and $P_{j2}$ at the two junctions
$F1$/$N$ and $N$/$F2$:
\begin{eqnarray} \label{eq:FNF11}
P_{j1} &=& P_{j1}^0 R_1\frac{R_N \coth(d/L_{sN}) +
R_{c2} + R_{F2} }{D_0} + P_{j2}^0 \frac{R_2 R_N}{D_0 \sinh(d/L_{sN})}, \\
\label{eq:FNF12} P_{j2} &=& P_{j2}^0 R_2 \frac{R_N \coth(d/L_{sN}) +
R_{c1} + R_{F1}}{D_0} + P_{j1}^0 \frac{R_1 R_N}{D_0
\sinh(d/L_{sN})}.
\end{eqnarray}
Here
\begin{equation}
D_0 = R_N^2 + (R_{c1} + R_{F1})(R_{c2}+R_{F2}) + R_N(R_{c1} + R_{F1}
+ R_{c2} + R_{F2}) \coth(d/L_{sN}),
\end{equation}
and $P_{j1}^0$ and $P_{j2}^0$ are the spin injection efficiencies of
the individual junctions giving by Eq. \ref{eq:SIE4}; similarly
$R_1$ and $R_2$ are the two effective junction resistances:
\begin{equation} \label{eq:FNF40}
R_1 = R_{F1} + R_{c1} + R_N, \quad R_2 = R_{F2} + R_{c2} + R_N.
\end{equation}

For a thick $N$ region, if $d \gg L_{sN}$, we recover the spin
injection efficiencies of the individual junctions: $P_{j1} \approx
P_{j1}^0$ and $P_{j2} \approx {P}_{j2}^0$, as expected for spin
uncoupled contacts. In the opposite limit of a thin $N$, if $d \ll
L_{sN}$,
\begin{eqnarray}
P_{j1}  = P_{j2} = \frac{{P}_{j1}^0 R_1 + {P}_{j2}^0 R_2}{R_{c1} +
R_{F1} + R_{c2} + R_{F2}}.
\end{eqnarray}
The spin injection efficiencies are a weighted mixture of the
efficiencies of the two individual junctions. Finally, for tunnel
contacts, such as $R_c \gg R_N$, $R_{F}$, and $R_N(L_s/a)$, we
recover the limit of independent junctions, $P_{j1} \approx
{P}_{j1}^0$ and $P_{j2} \approx {P}_{j2}^0$.

\paragraph{Nonequilibrium resistance.}
In order to find the value of the nonequilibrium resistance due to the
spin bottleneck, we need to find the increase of the quasichemical potential
$\Delta \mu$ across the junction:
\begin{equation}
{\rm emf}  = \mu_{F2} (\infty) - \mu_{F1} (-\infty) = j {\cal R} +
j\delta {\cal R},
\end{equation}
where
\begin{equation}
{\cal R} = {\cal R}_{F1} + {\cal R}_{c1} + {\cal R}_N + {\cal
R}_{c2} + {\cal R}_{F2},
\end{equation}
is the electrical resistance of the junction in the absence of spin
accumulation.

Let us apply Eq. \ref{eq:SCC9} to the five regions of the
$F$/$N$/$F$ junction: $F1$, $C1$, $N$, $C2$, and $F2$. In this
sequence, the regional increases of the quasichemical potential are
\begin{eqnarray}
\mu_{F1}(0)  -  \mu_{F1}(-\infty) & = & j {\cal R}_{F1} - P_{\sigma F1} \mu_{sF1}(0), \\
\mu_{N}(0)  -  \mu_{F1}(0) & = & j {\cal R}_{c1} - P_{\Sigma1} \Delta \mu_{s}(0), \\
\mu_{N}(d) -  \mu_{N}(0) & = & j {\cal R}_{N},  \\
\mu_{F2}(d)  -  \mu_{N}(d) & = & j {\cal R}_{c2} - P_{\Sigma2} \Delta \mu_{s}(d), \\
\mu_{F2}(\infty)  -  \mu_{F2}(d) & = & j {\cal R}_{F2} + P_{\sigma
F2} \mu_{sF2}(d).
\end{eqnarray}
We have used that $\mu_{sF1}(-\infty) = \mu_{sF2}(\infty) = 0$. Summing these equations
up we extract
\begin{equation}
j \delta {\cal R} = - P_{\sigma F1} \mu_{sF1}(0) - P_{\Sigma 1}
\Delta \mu_s(0) - P_{\Sigma 2} \Delta \mu_s(d) + P_{\sigma
F2}\mu_{sF2} (d).
\end{equation}
Expressing the spin chemical potentials in terms of the spin injection efficiencies
$P_{j1}$ and $P_{j2}$,
\begin{eqnarray}
\mu_{sF1}(0) & = & (P_{j1} - P_{\sigma F1}) j R_{F1}, \\
\Delta \mu_{s}(0) & = & (P_{j1} - P_{\Sigma1}) j R_{c1}, \\
\Delta \mu_{s}(d) & = & (P_{j2} - P_{\Sigma2}) j R_{c2}, \\
\mu_{sF2}(d) & = & (P_{\sigma F2} - P_{j2}) j R_{F2},
\end{eqnarray}
we find for the nonequilibrium resistance
\begin{equation}
\delta R = - P_{\sigma F1} (P_{j1} - P_{\sigma F1}) R_{F1} -
P_{\Sigma 1} (P_{j1} - P_{\Sigma 1}) R_{c1} - P_{\Sigma 2} (P_{j2} -
P_{\Sigma 2}) R_{c2} - P_{\sigma F2} (P_{j2} - P_{\sigma F2})
R_{F2}.
\end{equation}

\paragraph{Resistance difference $\Delta {\cal R}$.}  Denoting as
\begin{eqnarray}
\Delta P_{j} & \equiv & P_{j}^{\uparrow\downarrow} - P_{j}^{\uparrow \uparrow}, \\
\Delta^+ P_{j}&  \equiv&  P_{j}^{\uparrow\downarrow} +
P_{j}^{\uparrow \uparrow},
\end{eqnarray}
the difference and the sum of the spin injection efficiencies for
antiparallel and parallel magnetizations of the two ferromagnets, we
find (assuming that the spin efficiencies are positive, for example)
\begin{equation}
\Delta {\cal R} = -(R_{F1} P_{\sigma F1} + R_{c1} P_{\Sigma 1})
\Delta P_{j1} - (R_{F2} P_{\sigma F2} + R_{c1} P_{\Sigma 2})
\Delta^+ P_{j2}.
\end{equation}
From Eqs. \ref{eq:FNF11} and \ref{eq:FNF12} we find
\begin{eqnarray}
\Delta P_{j1} &=& - 2 \frac{R_2 R_N}{D_0\sinh(d/L_{sN})} {P}_{j2}^0, \\
\Delta^+ P_{j2} &=& - 2 \frac{R_1 R_N}{D_0\sinh(d/L_{sN})}
{P}_{j1}^0.
\end{eqnarray}
With that we finally get our desired result,
\begin{equation}
\Delta {\cal R} = \frac{4 R_1 R_2}{D_0 \sinh(d/L_{sN})} R_N
|{P}_{j1}^0 P_{j2}^0|.
\end{equation}
As expected, $\Delta {\cal R}$ vanishes exponentially if $d\gg
L_{sN}$, as the differences between parallel and antiparallel cases
diminish. On the other hand, for $d \ll L_{sN}$,
\begin{equation} \label{eq:FNF20}
\Delta {\cal R} \approx  \frac{4 R_1 R_2}{R_{F1}+ R_{c1} + R_{c2} +
R_{F2}} P_{j1}^0 P_{j2}^0.
\end{equation}

\paragraph{Transparent contacts.} Put $R_{c1} = R_{c2} = 0$ and consider the
interesting case of a thin $N$ layer, $d \ll L_{sN}$.  For simplicity assume the
same ferromagnets, $R_{F1} = R_{F2}$. Then
\begin{equation}
\delta {\cal R} \approx 2 R_F P_{\sigma F}^2.
\end{equation}

\vspace{0.4cm} \hspace{0.05\textwidth}
\begin{minipage}{0.9\textwidth}{\small {\bf Problem.} Analyze $\Delta {\cal R}$ in Eq. \ref{eq:FNF20} in
the conductivity mismatch regime, $R_N \gg R_c$, $R_F$.}
\end{minipage}
\vspace{0.4cm}

\paragraph{Tunnel contacts.} Suppose now that the most resistive regions are
the contacts and the $N$ region is thin, $d \ll L_{sN}$.
Assuming a symmetric junction, the nonequilibrium resistance
difference is
\begin{equation}
\Delta {\cal R} \approx \frac{2R_c P_\sigma^2}{1+ (R_c/R_N)
(d/2L_{sN})}.
\end{equation}
The spin accumulation detection by $\Delta R$ will be most sensitive
if
\begin{equation}
d R_c \ll  L_{sN} R_N,
\end{equation}
as then $\Delta {\cal R} \approx R_c$ and the resistance change is
maximized. Let us find the physical meaning of the above inequality
by invoking the definition of $R_N$, the diffusion length $L_{sN}$,
and the Einstein relation:
\begin{equation}
1 \gg \frac{d r_c}{r_N L_{sN}} = \frac{d}{L_{sN}^2} r_c \sigma_N =
\frac{d}{D_N \tau_{sN}} r_c e^2 g_N D_{N} \approx  e^2 (d g_N)
\frac{1}{\Sigma_c} \frac{1}{\tau_s}.
\end{equation}
Expressing the tunnel conductance $\Sigma_c$ through an effective tunneling probability
per unit time, $P_{\rm tunnel}$,
\begin{equation}
\Sigma_c = e (d g_N) P_{\rm tunnel},
\end{equation}
and introducing the dwell time
\begin{equation}
\tau_{\rm dwell} = 1/P_{\rm tunnel},
\end{equation}
we come to the conclusion that the spin accumulation detection in $F$/$N$/$F$
tunnel junctions is most efficient if
\begin{equation}
\tau_{\rm dwell} \ll \tau_{sN}.
\end{equation}
In words, the dwell time of the electrons in between the two tunnel
barriers (the average time the electron spends in the $N$ region)
must be much smaller than the spin relaxation time.

\vspace{0.4cm}
\hspace{0.05\textwidth}
\begin{minipage}{0.9\textwidth}{\small {\bf Problem.} {\it $N$/$F$/$N$ junction.}
 Calculate the spin efficiency $P_j$ and the nonequilibrium
resistance $\delta {\cal R}$ for a symmetric $N$/$F$/$N$ junction. \\
\noindent a) Show that in the limit of a thin $F$ layer ($d \ll L_{sF}$)
\begin{eqnarray}
P_j & = &  \frac{R_c P_\Sigma}{R_c + R_N}, \\
\delta {\cal R} & = &  \frac{2 R_c R_N P_\Sigma^2}{R_c + R_N}.
\end{eqnarray}
\\
\noindent b) Verify that in the limit of a thick $F$ layer ($d \gg
L_{sF}$) the spin injection efficiency $P_j$ reduces to its value
for a single $F$/$N$ junction, and that $\delta {\cal R}$ of an
$N$/$F$/$N$ junction is twice the nonequilibrium resistance of the
individual $F$/$N$ junctions.}
\end{minipage}
\vspace{0.4cm}

\section{Nonlocal spin-injection geometry: Johnson-Silsbee spin injection
experiment.}

In the $F$/$N$/$F$ junction studied in the previous section the
electric current flows through both contacts. As the current often
brings spurious effects from the point of view of spin detection,
especially in the presence of an external magnetic field (the Hall
effect or anisotropic magnetoresistance), it is important to
consider spin injection geometries in which the spin detection
circuit is open. We have already met one example of an open circuit
spin detection: the Silsbee-Johnson spin-charge coupling. This
scheme can be naturally extended to include a spin injection
contact, giving what is called a \emph{nonlocal spin-injection
geometry} (as the injection and detection circuits are independent)
or the \emph{Johnson-Silsbee spin injection experiment}, after the
original spin injection scheme.

\begin{figure} \label{fig:FNF_nonlocal1}
\epsfig{file=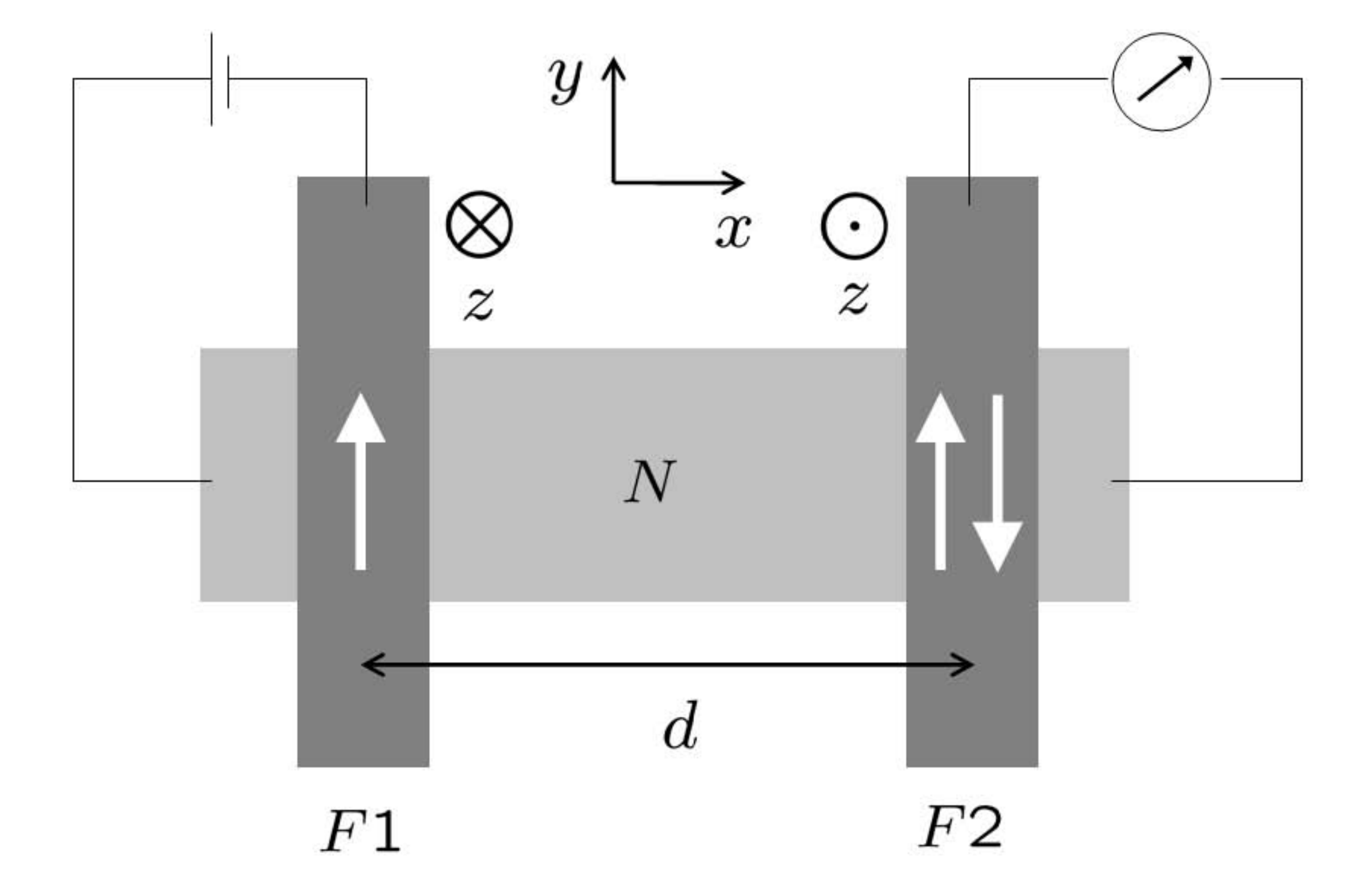,width=0.5\textwidth}
\caption{Nonlocal geometry for spin injection and detection. The
$F1$/$N$ circuit is closed, the $F2$/$N$ is open; the electrode $F1$
acts as a spin source, $F2$ as a spin drain. The axes labels are
indicated. The directions of the $z$ axis are opposite for the two
junctions.}
\end{figure}

Our goal is to answer the following question:

\vspace{0.2cm} {\it Suppose electric current drives spin injection
in the spin injection circuit $F1$/$N$
     as indicated in Fig. 7.
      What is the emf in the open $F2$/$N$ junction?}
\vspace{0.2cm}

The two ferromagnetic electrodes are on the top of a nonmagnetic
conductor, separated  by spin-polarizing contacts. Spin is
injected into $N$ from $F1$. While the electric current flows in the
closed circuit formed by $F1$/$N$, the spin current flows also
towards the spin detection circuit $N$/$F2$. The charge and spin
flows are indicated in Fig. 8. For spin the
contact $F1$/$N$ acts as a spin source, while $N$/$F2$ as a spin
sink. The source and sink will appear as special boundary conditions
for the spin transport in $N$. The axes labels are defined in Fig.
7.

We need to be a little careful with this geometry since in principle
we are now dealing with a two (if not three) dimensional problem. Nevertheless, the
problem can be decoupled to one-dimensional ones if we assume,
realistically, that the dimensions of the ferromagnetic electrodes
are much greater than the spin diffusion length in the ferromagnets.
Indeed, a representative $L_{sF}$ would be on the order of 10 nm or
so. In that case we can consider the spin current in $F1$ and $F2$
one-dimensional, along $z$. On the other hand, we assume that the
contact dimensions between $F$ and $N$, as well as the thickness of
$N$, are much smaller than the spin diffusion length in the
non-magnetic conductor, so that the spin current in $N$ can be
considered one-dimensional as well. Typically $L_{sN}$ would be more
than 1 $\mu$m. In most other cases one would need to set up a
two-dimensional drift-diffusion problem.\footnote{Suppose, for
example, that the thickness of $N$ would be much greater than
$L_{sN}$. Then the spin injected from $F1$ would diffuse not only
left and right, along $x$, but also down, along $z$, forming a
complicated diffusion profile. If the contact would be point-like,
the surface of an equal spin density would be a semisphere.}

With the above physical restrictions, the quantities labeled $N$
vary along $x$, while those of $F_1$ and $F_2$ along $z$, as
indicated in Figs. 7 and 8. For example, if we write
$\mu_{sF2}(0)$ we mean $\mu_{sF2}(z=0)$, the value of the spin
quasichemical potential in $F2$ at the place of contact with $N$.
Any variation of $\mu_{sF2}$ along $y$ or $x$ is insignificant,
occurring at the contact edges only.

We now apply the boundary conditions for the spin quasichemical
potentials $\mu_s$ at infinities:
\begin{equation}
\mu_{sN}(\pm \infty) = \mu_{sF1}(-\infty) = \mu_{sF2}(\infty) = 0.
\end{equation}
Let us consider each junction separately.

\begin{figure} \label{fig:FNF_nonlocal2}
\epsfig{file=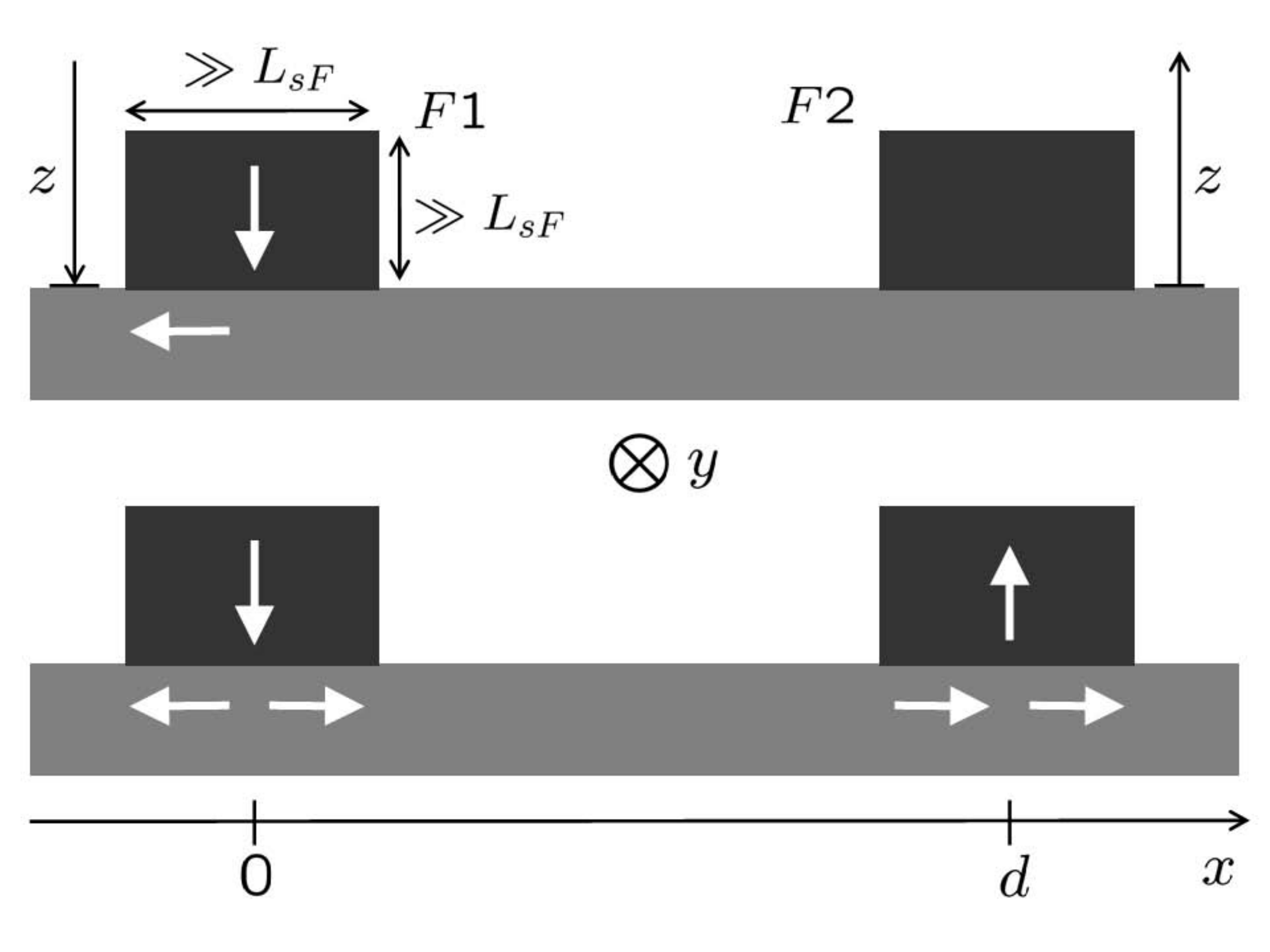,width=0.5\textwidth} \caption{Cross
view of the nonlocal geometry. The upper graph indicates the flow of
electrons, the lower graphs shows the flow of spins. From the point
of view of the spin flow in $N$, the injector plays a role of a spin
source, while the detector acts as a spin sink, taking away some of
the spin current.}
\end{figure}

\paragraph{Spin injector: $F1$/$N$ junction.} The distribution of
the spin currents is shown in Fig. 8. In $F1$
the spin accumulation has the profile
\begin{equation}
\mu_{sF1}(z) = \mu_{sF1}(0)e^{z/L_{sF2}},
\end{equation}
giving for the spin current at the contact, using Eq. \ref{eq:SC9}
\begin{equation}
j_{sF1}(0) = j P_{\sigma F1} + \frac{1}{R_{F1}} \mu_{sF1}(0).
\end{equation}
The spin current through the spin-polarizing contact $C1$ is
\begin{equation}
j_{sc1} = j P_{\Sigma 1} + \frac{1}{R_{c1}} \left [ \mu_{sN}(0) -
\mu_{sF1}(0) \right ].
\end{equation}
To obtain the spin current in $N$, we need to know the profile of
the spin quasichemical potential $\mu_{sN}$. Treating the chemical
potential at $x=0$ and $x=d$ as yet unknown, the profile is given by
Eq. \ref{eq:FNF9} for $x>0$. The profile in the whole $N$ region is
\begin{eqnarray}
\mu_{sN}(x\le 0) & = & \mu_{sN}(0) e^{x/L_{sN}}, \\
\mu_{sN} (0<x<d) & = & \mu_{sN}(d)
\frac{\sinh(x/L_{sN})}{\sinh(d/L_{sN})} - \mu_{sN}(0)
\frac{\sinh\left [(x-d)/L_{sN}\right ]}{\sinh(d/L_{sN})} , \\
\mu_{sN}(x\ge d) & = & \mu_{sN}(d) e^{-x/L_{sN}}.
\end{eqnarray}
The spin current is not continuous at $x=0$, due to the presence of
the spin source:
\begin{eqnarray}
j_{sN} (0+) & = & \frac{1}{R_N} \left [-\mu_{sN}(0) \coth\left
(d/L_{sN}\right ) + \frac{\mu_{sN}(d)}{\sinh\left (d/L_{sN}\right)}
\right ], \\
 j_{sN}(0-)& = & \frac{1}{R_N} \mu_{sN}(0).
\end{eqnarray}
The continuity of the spin current at the contact requires
that\footnote{Mind the sign convention: the reference current is
always along the positive axis.}
\begin{equation}
j_{sN}(0+) = j_{sN}(0-) + j_{sc1} = j_{sN}(0-) + j_{sF1}.
\end{equation}
Using the above algebraic system, we find
\begin{equation} \label{eq:nonlocal11}
\mu_{sN}(0)\left [\frac{R_N}{R_{c1} + R_{F1}} + \frac{\exp\left
(d/L_{sN}\right )}{ \sinh\left ( d/L_{sN} \right )} \right ] -
\mu_{sN}(d)\frac{1}{\sinh\left (d/L_{sN} \right)} = - jR_N
\frac{P_{\Sigma 1} R_{c1} + P_{\sigma F1} R_{F1}}{R_{c1} + R_{F1}}.
\end{equation}
The spin current in the spin injector contact is
\begin{equation}
j_{sF1} =  j \frac{P_{\Sigma 1} R_{c1} + P_{\sigma F 1} R_{F1} +
\mu_{sN}(0)/j}{R_{c1} + R_{F1}}.
\end{equation}
Finally, we define the spin injection efficiency for the spin
injector in the nonlocal geometry as
\begin{equation}
P_{j1} = \frac{j_{sN}(0+)}{j}.
\end{equation}
Only the spin current which gives rise to the spin accumulation at
the detector circuit is relevant.

\paragraph{Spin detector: $F2$/$N$ junction.} There is no electric
current flowing in the spin detector: $j=0$. The flow of spin is
indicated in Fig. 8. Similarly to the $F1$/$N$
junction, we obtain for the spin currents in the ferromagnet and the
contact,
\begin{eqnarray}
j_{sF2}(0) & = & - \frac{1}{R_{F2}} \mu_{sF2}(0), \label{eq:nonlocal22}\\
j_{sc2} & = & \frac{1}{R_{c2}} \left [\mu_{sF2}(0) - \mu_{sN}(d)
\right ] \label{eq:nonlocal23}.
\end{eqnarray}
The spin current along $x$ is again discontinuous at $x=d$, due to
the presence of the spin sink:
\begin{eqnarray}
j_{sN} (d-) & = & \frac{1}{R_N} \left [
-\frac{\mu_{sN}(0)}{\sinh\left (d/L_{sN}\right)} + \mu_{sN}(d)
\coth\left (d/L_{sN}\right )
\right ], \\
 j_{sN}(d+)& = & -\frac{1}{R_N} \mu_{sN}(d). \\
\end{eqnarray}
The continuity for the spin currents at $x=d$ gives
\begin{equation}
j_{sN}(d-) = j_{sN}(d+) + j_{sc2} = j_{sN}(d+) + j_{sF2}(0).
\end{equation}
Solving the above algebraic system yields
\begin{equation} \label{eq:nonlocal12}
\mu_{sN}(0) \frac{1}{\sinh\left (d/L_{sN} \right)}  - \mu_{sN}(d)
\left [\frac{R_N}{R_{c2} + R_{F2}} + \frac{\exp\left (d/L_{sN}\right
)}{ \sinh\left ( d/L_{sN} \right )} \right ]   = 0.
\end{equation}

Let us denote the spin current in the contact as $j_{s2}$:
\begin{equation}
j_{s2} \equiv j_{sc2} = j_{sF2}(0).
\end{equation}
We will need to know the value of this current to calculate the emf
at the detector circuit. We find that
\begin{equation} \label{eq:nonlocal31}
j_{s2} =- \frac{\mu_{sN}(d)}{R_{c2} + R_{F2}}.
\end{equation}

\paragraph{Spin quasichemical potentials.} Equations
\ref{eq:nonlocal11} and \ref{eq:nonlocal12} form a closed system,
allowing us to extract the spin quasichemical potentials at the two
contacts:
\begin{eqnarray}
\mu_{sN}(0) & = & - j \frac{R_N P_{j1}^0}{2} \left [ 1 -
\frac{R_N}{2R_2} \left ( 1 + e^{-2d/L_{sN}} \right ) \right ]
\frac{1}{\kappa}, \\ \label{eq:nonlocal43} \mu_{sN}(d) & = & -j
\frac{R_N P_{j1}^0}{2} \left (\frac{R_{c2} + R_{F2}}{R_2} \right )
\frac{e^{-d/L_{sN}}}{\kappa},
\end{eqnarray}
 where
\begin{equation}
\kappa = \frac{R_N^2}{4 R_1 R_2} \left [\left (1+ 2\frac{R_{c1} +
R_{F1}}{R_N} \right ) \left (1 + 2 \frac{R_{c2} + R_{F2}}{R_N}
\right ) - e^{-2d/L_{sN}} \right ].
\end{equation}
Recall that $R_1$ and $R_2$ are the total effective resistances of
the two junctions; see Eqs. \ref{eq:FNF40}. Similarly, $P_{j1}^0$
and $P_{j2}^0$ are the spin injection efficiencies of the individual
$F$/$N$ junctions, given by Eq. \ref{eq:SIE4}.

\vspace{0.4cm} \hspace{0.05\textwidth}
\begin{minipage}{0.9\textwidth}{\small {\bf Problem.} Calculate the
spin injection efficiency $P_{j1}$ of the spin injection circuit. What do you get in the
limit of  $d \gg L_{sN}$? Does the result agree with that for an
isolated $F$/$N$ junction studied earlier?  }
\end{minipage}
\vspace{0.4cm}

\paragraph{emf in the detector circuit.} Due to the presence of a
nonequilibrium spin in the detector circuit, an emf will develop
there. We can obtain it as the increase of the quasichemical
potential from the far end of the $F2$ to the far right of the $N$
region, as shown in Fig. 7. Since the spin
flow in $F2$ is confined to the distance of order $L_{sF2}$ from
$z=0$, the quasichemical potential $\mu_{sF2}$ far away from the
contact, at $y\to \pm \infty$ (we are mixing the third dimension
here!) will be the same as that at the contact itself, $y \approx 0$,
but at $z=\infty$:\footnote{Since $j=0$, we have that $\nabla \mu = -
P_\sigma \nabla \mu_s$. Integrating this equation in the $(y,z)$
plane of $F_2$ we get
\begin{equation}\mu_{F2}(y,z) = \mu_{F2}(y_0, z_0) - P_{\sigma F2}
\left [\mu_{sF2}(y,z) - \mu_{sF2} (y_0,z_0) \right], \end{equation}
where $(y_0, z_0)$ is a reference point. Choosing $y_0$ from the
contact region and letting $z_0 \to \infty$, we get that
$\mu_{F2}(\infty, z) \approx \mu_{F2}(y_0,\infty)$, since the spin
accumulation vanishes both at $y \to \infty$ and $z_0 \to \infty$.
The far ends of the $F2$ electrodes ($y_0 \to \infty$) are thus
equipotential with the $z=\infty$ points in the contact region.}
\begin{equation}
{\rm emf} = \mu_N(\infty) - \mu_{F2}(\infty).
\end{equation}
Since $j=0$ in the $F2$/$N$
junction, we can write,
\begin{eqnarray}
\mu_{F2}(0)  - \mu_{F2}(\infty) &=& -P_{\sigma F2} \mu_{sF2}(0), \\
\mu_N(d) - \mu_{F2}(0) & = & - P_{\Sigma 2} \left [\mu_{sN}(d) - \mu_{sF2}(0) \right ], \\
\mu_N(\infty) - \mu_N(d) &=& 0.
\end{eqnarray}
Summing these equations up we get, after substituting for the spin
quasichemical potentials Eqs. \ref{eq:nonlocal22} and
\ref{eq:nonlocal23},
\begin{equation}
{\rm emf} = (R_{c2} P_{\Sigma2} + R_{F2} P_{\sigma 2}) j_{s2}.
\end{equation}
This is just another realization of Silsbee-Johnson spin-charge
coupling: An electromotive force develops due to the presence of a
spin current in a spin-polarized contact or a ferromagnetic
conductor. Due to charge neutrality this emf can be detected as a
voltage drop.

Substituting for $j_{s2}$ using Eq. \ref{eq:nonlocal31} and using
Eq. \ref{eq:nonlocal43} for $\mu_{sN}(d)$, the emf can be readily
obtained:
\begin{equation} \label{eq:NLSIG10}
{\rm emf} =   j \frac {R_N}{2} {P}_{j1}^0 {P}_{j2}^0
\frac{e^{-d/L_{sN}}}{\kappa}.
\end{equation}
The emf is in general positive for parallel and negative for
antiparallel magnetization orientations.

Often what is detected is the \emph{nonlocal resistance},
\begin{equation}
{\cal R}_{\rm nl} = \frac{{\rm emf}}{j} =  \frac{R_N}{2} {P}_{j1}^0
{P}_{j2}^0 \frac{e^{-d/L_{sN}}}{\kappa},
\end{equation}
or the corresponding difference in the nonlocal resistance for
parallel and antiparallel orientations of the magnetizations of $F1$
and $F2$:
\begin{equation}
\Delta {\cal R}_{\rm nl} = {\cal R}_{\rm nl}^{\uparrow \uparrow} -
{\cal R}_{\rm nl}^{\uparrow \uparrow} = 2 |{\cal R}_{\rm nl}|.
\end{equation}

\paragraph{Tunnel contacts.} For tunnel contacts we find $\kappa \approx 1$ and
\begin{equation}
{\rm emf} = j \frac{R_N}{2} P_{\Sigma 1} P_{\Sigma 2} e^{-d/L_{sN}},
\end{equation}
as one would expect. The factor of ``1/2'' appears due the geometry
of the spin injector: only half of the injected spin current in the
$F1$/$N$ junction flows towards the $F2$/$N$ junction. The other
half flows towards $x\to -\infty$.

\paragraph{Transparent contacts.} The most general expression for transparent
contacts is the same as Eq. \ref{eq:NLSIG10}, with
$R_{c1}=R_{c2}=0$. In the conductivity mismatch regime, for $R_N \gg
R_{F1}, R_{F2}$, the emf simplifies to
\begin{equation}
{\rm emf} = 2 j R_N P_{\sigma F1} P_{\sigma F2} \left (\frac{R_{F1}
R_{F2}}{R_N^2} \right ) \frac{e^{-d/L_{sN}}}{1-e^{-2d/L_{sN}}}.
\end{equation}
The conductivity mismatch limits the spin injection/detection in the
nonlocal geometry.

\vspace{0.4cm} \hspace{0.05\textwidth}
\begin{minipage}{0.9\textwidth}{\small {\bf Problem.} {\it Tunnel/transparent contacts.}
Calculate emf for the mixed case of tunnel and transparent contacts
of the nonlocal spin injection geometry.}
\end{minipage}
\vspace{0.4cm}

\bibliography{../../../references_master}

\end{document}